\documentclass[a4paper,accepted=2025-01-29]{quantumarticle}
\pdfoutput=1

\usepackage{notes2bib}
\usepackage{gensymb}
\usepackage{amsthm}
\usepackage{amsmath}
\usepackage{color}
\usepackage{bm}
\usepackage{amsmath,amssymb}
\usepackage{amsfonts}
\usepackage{dsfont}
\usepackage{graphicx}
\usepackage{float}
\usepackage{subfigure}
\usepackage{verbatim}
\usepackage{amsfonts}
\usepackage{bbold}
\usepackage{dcolumn}
\usepackage{bm}
\usepackage[dvipsnames]{xcolor}
\usepackage{listings}
\definecolor{blueprl}{RGB}{46,48,146}
\usepackage[pdftex,colorlinks=true,urlcolor=blueprl,citecolor=blueprl,linkcolor=blueprl]{hyperref}
\usepackage{mathrsfs}
\usepackage{mathtools}
\usepackage{color}
\usepackage[dvipsnames]{xcolor}

\usepackage{braket}

\usepackage{soul}

\newcommand{\dd}{\dagger}

\usepackage{mathtools}
\usepackage{dsfont}
\usepackage[mathscr]{euscript}
\usepackage{amsmath}
\usepackage{graphicx}
\usepackage{dcolumn}
\usepackage{bm}
\usepackage{epsfig}
\usepackage{amssymb,latexsym,mathrsfs}
\usepackage{graphicx}
\usepackage{color}
\usepackage{hyperref}
\usepackage{float}
\usepackage{diagbox}
  \usepackage{cancel}
  \usepackage[normalem]{ulem}

\definecolor{vividviolet}{rgb}{0.62, 0.0, 1.0}
\definecolor{amaranth}{rgb}{0.9, 0.17, 0.31}
\definecolor{palatinateblue}{rgb}{0.15, 0.23, 0.89}
\definecolor{brightpink}{rgb}{1.0, 0.0, 0.5}
\definecolor{cornflowerblue}{rgb}{0.39, 0.58, 0.93}
\definecolor{deepcarminepink}{rgb}{0.94, 0.19, 0.22}
\definecolor{radicalred}{rgb}{1.0, 0.21, 0.37}
\definecolor{blueblue}{RGB}{21,47,181}
\definecolor{greengreen}{RGB}{65,166,16}
\newcommand{\ranglem}{|0_M\rangle }

\usepackage{xtab,afterpage}
\usepackage{hyperref}
\usepackage{setspace,calc}

\allowdisplaybreaks

\newcommand{\be}{\begin{equation}}
\newcommand{\ee}{\end{equation}}
\newcommand{\bs}{\begin{split}} 
\newcommand{\bea}{\begin{eqnarray}}
\newcommand{\eea}{\end{eqnarray}}
\newcommand{\infint}{\int_{-\infty }^{\infty }}
\newcommand{\vt}{\vphantom{\frac{1}{2}}}

\newcommand{\non}{\nonumber} 
 
\newcommand{\p}{\partial} 
\newcommand{\D}{\mathrm{d}}
\newsavebox{\myhbar}

\begin{document}

\title{Superpositions of thermalisations in relativistic quantum field theory}
\author{Joshua Foo}
\email{jfoo@uwaterloo.ca}
\affiliation{Centre for Quantum Computation \& Communication Technology, School of Mathematics \& Physics, The University of Queensland, St.~Lucia, Queensland, 4072, Australia}
\affiliation{Department of Physics, Stevens Institute of Technology, Castle Point Terrace, Hoboken, New Jersey 07030, U.S.A.}
\affiliation{Department of Physics and Astronomy, University of Waterloo, Waterloo, Ontario, Canada, N2L 3G1}
\author{Magdalena Zych}
\email{magdalena.zych@fysik.su.se}
\affiliation{Centre for Engineered Quantum Systems, School of Mathematics and Physics, The University of Queensland, St. Lucia, Queensland, 4072, Australia}
\affiliation{Department of Physics, Stockholm University, AlbaNova University Center, SE-106 91 Stockholm, Sweden}

\begin{abstract}
Recent results in relativistic quantum information and quantum thermodynamics have independently shown that in the quantum regime, a system may fail to thermalise when subject to quantum-controlled application of the same, single thermalisation channel. For example, an accelerating system with fixed proper acceleration is known to thermalise to an acceleration-dependent temperature, known as the Unruh temperature. However, the same system in a superposition of spatially translated trajectories that share the same proper acceleration fails to thermalise. Here, we provide an explanation of these results using the framework of quantum field theory in relativistic noninertial reference frames. We show how a probe that accelerates in a superposition of spatial translations interacts with incommensurate sets of field modes. In special cases where the modes are orthogonal (for example, when the Rindler wedges are translated in a direction orthogonal to the plane of motion), thermalisation does indeed result, corroborating the here provided explanation. We then discuss how this description relates to an information-theoretic approach aimed at studying quantum aspects of temperature through quantum-controlled thermalisations. The present work draws a connection between research in quantum information, relativistic physics, and quantum thermodynamics, in particular showing that relativistic quantum effects can provide a natural realisation of quantum thermodynamical scenarios. 
\end{abstract}

\maketitle 

\section{Introduction}
Historically, quantum thermodynamics has sought to establish the laws of classical thermodynamics from quantum mechanics \cite{kosloffe15062100,andersdoi:10.1080/00107514.2016.1201896}.\ A key development in the field was the Markovian master equation proposed by Lindblad and Gorini-Kossakowski-Sudarshan \cite{Lindblad1976,gorinidoi:10.1063/1.522979}, which supplied a framework for analysing the quantum dynamics of systems interacting with an environment \cite{breuer2002theory}. Apart from its fundamental significance, quantum thermodynamics is closely tied to applied research fields such as ultracold atomic systems \cite{gardiner2015quantum} and quantum information processing \cite{calderbankPhysRevA.54.1098}, both widely regarded as forming the basis of emerging quantum technologies. 

Beyond this, quantum thermodynamics motivates a related set of fundamental questions, namely how one characterises classical, macroscopic phenomena such as energy, work, and temperature when they are subject to quantum indeterminacy. For example, significant attention has been devoted to understanding the effect of applying quantum channels in coherent superpositions \cite{howl2022quantum}, including that of causal order \cite{procopioPhysRevA.101.012346,Abbott2020communication,gaoPhysRevLett.124.030502,Paunkovic2020causalorders,dahlstenPhysRevLett.129.230604,BAN2021127381}, with demonstrated advantages in heat cycle performances \cite{felcePhysRevLett.125.070603,niePhysRevLett.129.100603}, quantum computation \cite{Chiribella_2021}, and other thermodynamical processes \cite{CHAPEAUBLONDEAU2022128300,guhaPhysRevA.102.032215,simonovPhysRevA.105.032217}. 

Here, we are motivated by the problem of characterising the quantum aspects of temperature, which has recently emerged as a problem of interest in different contexts. Quantum theory allows systems to exist in delocalised superpositions, and thus for scenarios in which such systems interact with different thermal environments at different temperatures. A concrete example at the intersection of quantum thermodynamics and general relativity is the coupling of matter to thermal radiation emitted from event horizons, either due to rapid accelerations (the Unruh effect) or black holes (the Hawking effect) \cite{unruh1976notes,unruh1984happens,hawking1974black}. Recent results have shown that two-level systems (e.g.\ Unruh deWitt detectors) travelling in a superposition of proper accelerations, or situated outside a mass-superposed black hole do not exhibit the usual Planckian response at the Unruh or Hawking temperatures associated with either of the worldlines in superposition, or even some combination of those temperatures \cite{foo2020unruhdewitt,fooPhysRevResearch.3.043056,Foo_2021schrodingers,dimic2017simulating,barbadoPhysRevD.102.045002,fooPhysRevLett.129.181301}.  

More counter-intuitive results have demonstrated that even for superpositions in which each amplitude is associated with the same temperature (e.g.\ superpositions of Rindler trajectories with equal proper accelerations but translated in the direction of motion by a constant offset), the system still does not thermalise to the Unruh temperature associated with that acceleration \cite{foo2020unruhdewitt}. Recently, a related result was obtained in a general setting using a Kraus operator approach to quantum-controlled superpositions of channels acting on a probe particle \cite{wood2021https://doi.org/10.48550/arxiv.2112.07860}. This paper by Wood et.\ al.\ among others demonstrated lack of thermalisation when the probe interacted with a bath in a superposition of different purified states, but each yielding the same reduced thermal state of the bath. 

In this article, we elucidate a formal connection between the abstract framework of Wood et.\ al.\ and the scenarios involving relativistic accelerating systems interacting with a quantum field~\cite{crispino_2008RevModPhys.80.787}.\ Specifically, we combine ideas and techniques from quantum thermodynamics and relativistic quantum physics via the framework of quantum field theory in noninertial reference frames. We use well-understood reference frames associated with accelerated and strictly localised observers (Rindler wedges and spacetime diamonds respectively) and show how an observer in a superposition of these reference frames will detect a nonthermal particle spectrum in the Minkowski vacuum. This nonthermalisation originates from the fact that the observer couples to two incommensurate sets of field modes, characterised by two inequivalent Bogolibov decompositions with respect to the global Minkowski modes. Thus, when the wedges (or diamonds) are not fully orthogonal, complex correlations exist between the modes confined within \cite{sucommPhysRevD.90.084022}. On the other hand, if the observer or probe is placed in a superposition of trajectories enabling them to interact with the same set of modes, or a fully orthogonal set of modes, thermalisation does indeed result. This latter result has gone unnoticed in prior works (e.g.\ \cite{felcePhysRevLett.125.070603,wood2021https://doi.org/10.48550/arxiv.2112.07860}), since the thermal baths are modelled with a generic Kraus decomposition that ignores the details of this mode structure. Our focus on this underlying mode structure is a unique strength of our present analysis, since it allows us to characterise thermalisation or lack thereof in the case of superpositions via standard operational notions in quantum field theory (e.g.\ the frequency spectrum of detected particles or the Kubo-Martin-Schwinger condition).

More generally, our results draw a connection between the disciplines of relativistic physics, quantum theory, and thermodynamics by proposing concrete scenarios where phenomena from all three are relevant. This is in the spirit of pioneers like Bekenstein \cite{bekensteinPhysRevD.7.2333}, Bardeen, Carter, and Hawking \cite{Bardeen1973} among others, who applied the principles of quantum information and thermodynamics to black holes, thereby originating the field of black hole thermodynamics. Similarly, exploring the notions of temperature and other thermodynamical phenomena when the motion of observers or detectors are quantised degrees of freedom is of foundational interest. For example, the framework provided here motivates further questions concerning the thermodynamics of quantum reference frames (in our case, the observer or probe defines a reference frame ``in superposition'') \cite{giacomini2019}, which to our knowledge has not hitherto been studied. The present work is partly motivated by such questions, more specifically the notion of temperature when the ``thermometer'' is a relativistic quantum system and the ``bath'' is a relativistic quantum field.

This article is structured as follows. We begin with a review of the Rindler and diamond coordinates, before reviewing a derivation of the Unruh effect in these respective coordinates via Bogoliubov transformations. We then proceed with our main results, in which we consider a superposition of spatially translated Rindler wedges (diamonds), and show that the inequivalent decomposition of the modes--with respect to the original Minkowski coordinates--leads to a nonthermal particle distribution. We study some examples--including previously unexplained results utilising Unruh-deWitt detectors in cosmological and black hole spacetimes--in which an observer in a superposition of trajectories does detect a thermal particle distribution, before concluding with some final remarks. 

Throughout, we utilise natural units $c = \hslash = k_B = 1$.

\section{Rindler and Diamond Coordinates}

\subsection{Rindler Coordinates}
Let us first review the Rindler coordinates, describing uniformly accelerated observers in Minkowski spacetime. The worldline of a uniformly accelerated observer in the right Rindler wedge of Minkowski spacetime, for which we use coordinates $(t,x,y,z)$,  can be described in terms of the Rindler coordinates $(\tau, \xi,y,z)$ \cite{crispino_2008RevModPhys.80.787},
\begin{align}
    t &= a^{-1} e^{a\xi} \sinh ( a  \tau ) ,
    \vt  
    \\
    x &= a^{-1} e^{a\xi} \cosh ( a \tau ) , 
    \vt 
    \label{eq3.12}
\end{align}
with all other coordinates constant, where $\tau$ and $a$ are the proper time and acceleration of the observer. In these coordinates, the line element takes the form, 
\begin{align}
    \D s^2 &= - e^{2a \xi} ( \D \tau^2 - \D \xi^2 ) + \D y^2 + \D z^2 .
    \vt 
\end{align}
A corresponding set of coordinates $(\bar{\tau}, \bar{\xi}, y,z)$ describes the left Rindler wedge:
\begin{align}
    t &= - a^{-1}e^{a\bar{\xi}} \sinh( a \bar{\tau} ) ,
    \vt 
    \\
    x &= - a^{-1}e^{a\bar{\xi}} \cosh(a\bar{\tau} ) ,
    \vt 
\end{align}
with line element 
\begin{align}
    \D s^2 &= -e^{2a \bar{\xi}} ( \D \bar{\tau}^2 - \D \bar{\xi}^2 ) + \D y^2 + \D z^2 . 
    \vt 
\end{align}
Coordinates can also be identified for the future and past lightcones respectively \cite{olson2011entanglement}.\ The entire Minkowski spacetime is thus covered by these four Rindler wedges/lightcones, separated by the null hypersurfaces $t = \pm x$. These can be interpreted as a kind of event horizon, in that uniformly accelerated observers moving on trajectories of constant $\xi$, $\bar{\xi}$ will remain spacelike separated from the events on the other side of the corresponding horizon. The Unruh effect, described in detail below, can be explained with reference to such a horizon as follows: modes localised behind the horizon (those localised in the left wedge) are inaccessible to systems following any of the accelerated trajectories in the right wedge; upon tracing out these modes one finds that the state of the field in the right wedge is thermal--which in turn is due to the fact that the Minkowski vacuum is an entangled state of the modes in the right and left Rindler wedges~\cite{suPhysRevX.9.011007,fooPhysRevD.101.085006}.  

\begin{figure}[h]
    \centering
    \includegraphics[width=0.75\linewidth]{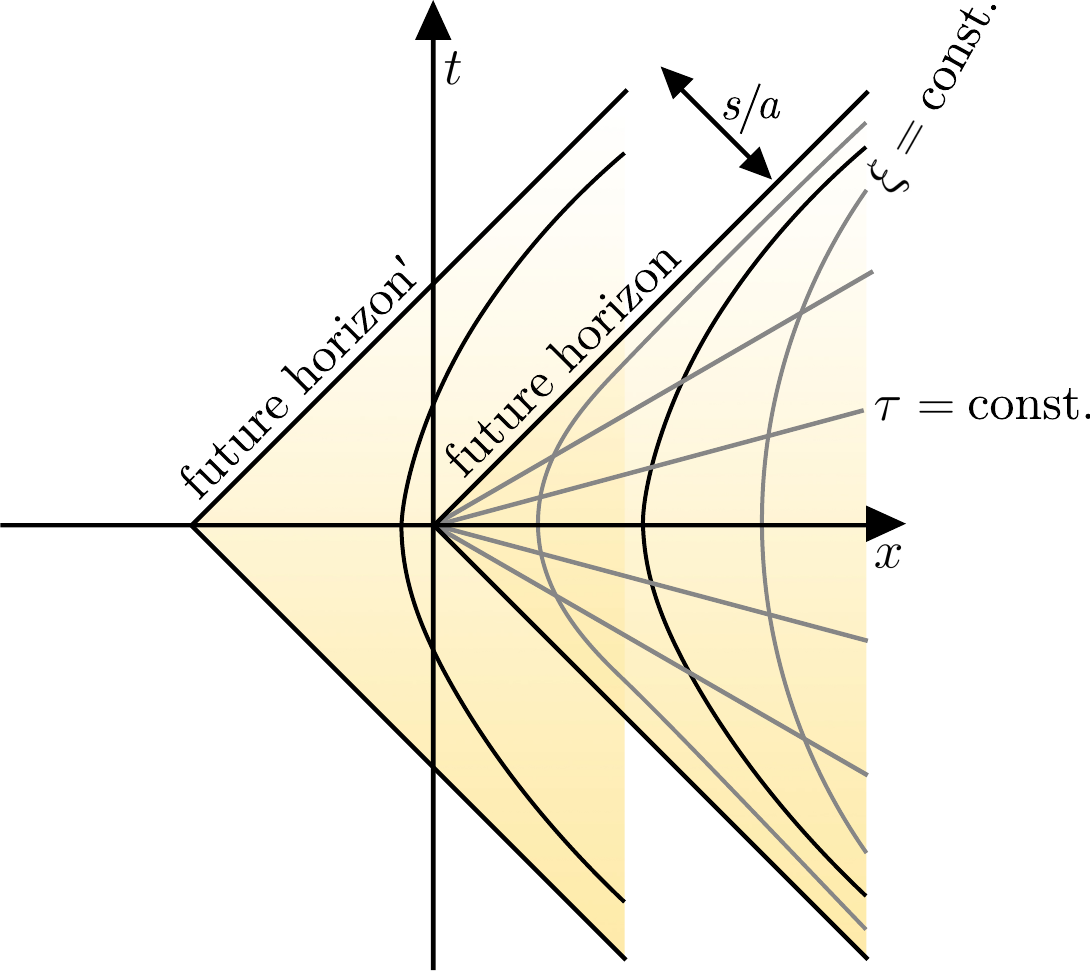}
    \caption[Schematic diagram of the Rindler coordinates]{Schematic diagram of the Rindler coordinates. The black lines represent hypersurfaces of constant $\tau$, while the grey hyperbolas are surfaces of constant $\xi$. The second Rindler wedge is shifted by $s/a$ in the null coordinate $V$, according to an inertial observer.}
    \label{fig:rindlerspace}
\end{figure}

Figure \ref{fig:rindlerspace} illustrates an accelerated trajectory with constant $\xi$ in the right Rindler wedge, as well as another wedge shifted in the null coordinate $V$ by the distance (according to an inertial observer) $s/a$. Below, we study this particular scenario with the accelerated observer travelling in a superposition of these spatially translated worldlines. 

\subsection{Diamond Coordinates}
The Rindler coodinates define unbounded regions of spacetime that partition Minkowski spacetime into disjoint regions. Another set of coordinates that possesses a similar feature are those parametrising so-called spacetime diamonds. Interest in this coordinate system increased following Martinetti and Rovelli's application of the thermal time hypothesis to derive the diamond temperature for finite-lifetime observers, a generalisation of the Unruh effect for such observers \cite{martinetti2003diamond,ida2013modular}. More recently it has been proposed as a way of witnessing vacuum entanglement \cite{su2016spacetime,Foo_2020generating,chakPhysRevD.106.045027} and has attracted interest due to its relevance for conformal quantum mechanics \cite{camblongPhysRevD.110.124043,camblong10.1063/5.0150349} and its relationship with the de Sitter universe \cite{jacobson10.21468/SciPostPhys.7.6.079,goodPhysRevD.102.045020,GIBBONS2007317,berthierePhysRevD.92.064036,Arzano2020}.  

\begin{figure}[h]
    \centering
    \includegraphics[width=0.95\linewidth]{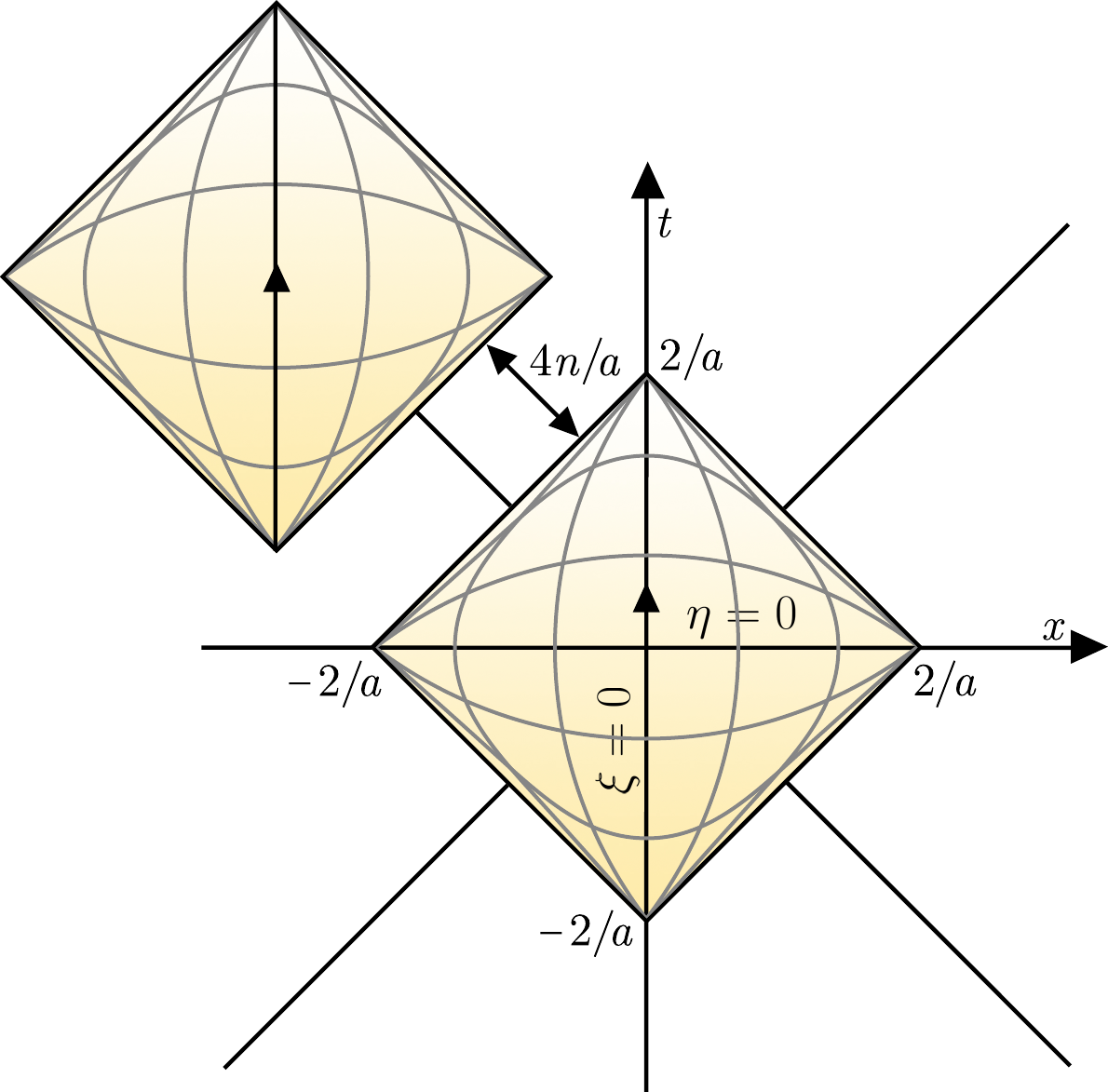}
    \caption[Schematic diagram of the diamond]{Schematic diagram of the diamond with localisation parametrised by $a$. The grey lines represent hypersurfaces of constant $\eta$, $\xi$. The $n$th diamond is shifted by $4n/a$ with respect to the zeroth diamond. }
    \label{fig:diamond illustration}
\end{figure}

A static observer (who stays at $\mathbf{r}=(x,y,z)= 0$) with a finite lifetime lives in a diamond, defined as the overlapping region between the future and past lightcones at their birth and death respectively \cite{su2016spacetime,ida2013modular,martinetti2003diamond}. We refer to the diamond centred on the origin of Minkowski coordinates as the zeroth diamond, while those translated by $4n/a$ in the null coordinate $V = t + x$ (i.e.\ when specialising to (1+1)-dimensions, below) are referred to as the $n$th diamond(s), see Fig.\ \ref{fig:diamond illustration}. The $n$th and $(n+1)$th diamonds share a common boundary. Each diamond satisfies $|t| + |\mathbf{r}| < 2/a$ where $\mathcal{T} = 4/a$ is the lifetime of the observer. There exists a conformal transformation that maps the diamond to a Rindler wedge. Using again  $(t,x,y,z)$ for the Minkowski coordinates and defining $(t',x',y',z')$ as the conformal coordinates, the conformal transformation is defined as \cite{su2016spacetime}
\begin{align}
    \frac{at'}{2} &= \frac{at}{f_-(t,\mathbf{r};a)}
    \vphantom{\frac{1 + (at/2)^2 - (ar/2)^2}{f_-(t,\mathbf{r};a)}} ,
    \vphantom{\frac{ax'}{2} = \frac{1 + (at/2)^2 - (ar/2)^2}{f_-(t,\mathbf{r};a)} ,}
    \non 
\\
    \frac{ax'}{2} &= \frac{1 + (at/2)^2 - (ar/2)^2}{f_-(t,\mathbf{r};a)} ,
    \non 
\\
    \frac{ay'}{2} &= \frac{ay}{f_-(t,\mathbf{r};a)} 
    \vphantom{\frac{1 + (at/2)^2 - (ar/2)^2}{f_-(t,\mathbf{r};a)}} ,
    \vphantom{\frac{ax'}{2} = \frac{1 + (at/2)^2 - (ar/2)^2}{f_-(t,\mathbf{r};a)} ,}
    \non 
\\
    \frac{az'}{2} &= \frac{az}{f_-(t,\mathbf{r};a)} 
    \vphantom{\frac{1 + (at/2)^2 - (ar/2)^2}{f_-(t,\mathbf{r};a)}} ,
    \vphantom{\frac{ax'}{2} = \frac{1 + (at/2)^2 - (ar/2)^2}{f_-(t,\mathbf{r};a)} ,}
\end{align}
where $f_- (t,\mathbf{r};a) = 1 - (at/2)^2 + (ar/2)^2 - ax$ and $r = \sqrt{x^2 + y^2 + z^2}$. The line element in the conformal coordinates reads
\begin{align}
    \D s^2 &= \frac{4}{f_+^2(t',\mathbf{r}';a)} (- \D t'^2 + \D x'^2 + \D y'^2 + \D z'^2),
    \vt 
\end{align}
where $f_+(t,\mathbf{r};a) = 1 - (at/2)^2 +(ar/2)^2 + ax$, which is consistent with the assumption that this is a conformal transformation. 

To describe spacetime events and field modes within the diamond, we introduce diamond coordinates ($\eta, \xi,\zeta,\rho)$, which are related to the Minkowski coordinates via the transformation \cite{su2016spacetime}
\begin{align}
    \eta &= a^{-1} \tanh^{-1} \bigg\{ \frac{at}{1 + (at/2)^2 - (ar/2)^2} \bigg\} \vphantom{\frac{\sqrt{(1)^2}}{(1)^2}}, 
    \non \\
    \xi &= a^{-1} \ln \bigg\{ \frac{\sqrt{(1 + (at/2)^2 - (ar/2)^2 )^2 - (at)^2}}{f_+(t,x,y,z;a)} \bigg\} ,
    \non \\
    \zeta &= \frac{2y}{f_+(t,x,y,z;a)}  \vphantom{\frac{\sqrt{(1)^2}}{(1)^2}} ,
    \non \\
    \rho &= \frac{2z}{f_+(t,x,y,z;a)} \vphantom{\frac{\sqrt{(1)^2}}{(1)^2}} .
\end{align}
It can be shown that $\zeta = \rho = 0$, $\xi = \mathrm{const.}$ corresponds to the worldline of a uniformly accelerated observer with acceleration $(a/2)|\sinh(a\xi)|$ according to an inertial observer \cite{su2016spacetime}. The case we are most interested in is $\zeta = \rho = \xi = 0$, which is the worldline of a static observer inside the diamond. On this worldline, the Minkowski time is related to the conformal time by $t = (2/a) \tanh(a\eta/2)$, which means that the diamond clock ticks at the same rate as an inertial clock at $\eta = 0$, while the former ticks much faster than the latter when $\eta \to \pm \infty$. In Fig.\ \ref{fig:diamond illustration}, we have illustrated a static worldline within the zeroth diamond and its spatially shifted counterpart,  within the $n$th diamond.

\section{Unruh Effect in Rindler and Diamond Reference Frames}
In this section, we review some preliminary derivations of the Unruh effect in the Rindler wedge and spacetime diamond, using Bogoliubov transformations of the modes with respect to the global Minkowski modes. For simplicity, we focus on a dimensionally reduced (1+1)-dimensional (massless) scalar field theory, which captures the essential physics while remaining calculationally simpler. 

\subsection{Unruh Effect in Rindler Coordinates}
Let us first consider a massless scalar field $\hat{\Phi}$ satisfying the Klein-Gordon equation $\Box \hat{\Phi} = 0$, which can be expanded in the complete orthonormal basis of Minkowski plane waves, 
\begin{align}
    \hat{\Phi}(U,V) &= \int\D k \: \Big( u_{kl}(V) \hat{a}_{kl} + u_{kr}(U) \hat{a}_{kr} + \mathrm{H.c} \Big),
\end{align}
where we have introduced lightcone coordinates $V = t + x$, $U = t - x$, $\mathrm{H.c}$ denotes the Hermitian conjugate, $\hat{a}_{kl}$, $\hat{a}_{kr}$ are annihilation operators of the left- and right-moving Minkowski modes with frequency $k$ respectively, and 
\begin{align}
    u_{kl}(V) &= \frac{1}{\sqrt{4\pi k}} e^{-ikV} ,
    \\
    u_{kr}(U) &= \frac{1}{\sqrt{4\pi k}} e^{-ikU} ,
\end{align}
are Minkowski plane wave mode functions.\ In (1+1)-dimensions, the left- and right-moving modes decouple, so one can separate the left- and right-moving sectors of the field as follows:
\begin{align}
    \hat{\Phi}(U) &= \int\D k \: \Big( u_{kr}(U) \hat{a}_{kr} + \mathrm{H.c} \Big) 
    , 
    \\
    \hat{\Phi}(V) &= \int\D k \: \Big( u_{kl}(V) \hat{a}_{kl} + \mathrm{H.c} \Big) .
\end{align}
Henceforth, we shall consider the left-moving modes only for brevity. In Rindler coordinates, the Klein-Gordon equation reads, 
\begin{align}
    \left( - \frac{\p^2}{\p \tau^2} + \frac{\p^2}{\p \xi^2} \right) \hat{\Phi} &= 0 ,
\end{align}
for which the field can be expanded in the Rindler modes, 
\begin{align}
    \hat{\Phi}(V) &= \int\D \omega \Big( g_{\omega}^R (V) \hat{b}_{\omega}^R + g_{\omega}^L(V) \hat{b}_{\omega}^L  + \mathrm{H.c} \Big) ,
\end{align}
where $\hat{b}_{\omega}^R$, $\hat{b}_{\omega}^L$ are single-frequency Rindler annihilation operators for the modes with frequency $\omega$ localised to the right and left wedges respectively, and the normalised mode solutions in terms of the Rindler frequency $\omega$ are given by:
\begin{align}
    g_{\omega}^R (\tau, \xi) &= \frac{1}{\sqrt{4\pi\omega}} \begin{cases} e^{-i\omega v} & x > | t | , 
    \\
    0 & x < - | t| , 
    \end{cases} 
    \\
    g_{\omega}^L (\bar{\tau}, \bar{\xi} ) &= \frac{1}{\sqrt{4\pi \omega}}
    \begin{cases}
    0 & x > | t|  , 
    \\
    e^{-i\omega \bar{v}} & x < - | t |,
    \end{cases} 
\end{align}
where $v = \tau + \xi$, $\bar{v} = - ( \bar{\tau } + \bar{\xi})$. The Rindler lightcone coordinates are related to the Minkowski coordinates by $(aV) = e^{av}$ and $(-aV) = e^{a\bar{v}}$. 

A Bogoliubov transformation relates the single-frequency Rindler operators with the single-frequency Minkowski operators, as follows:
    \begin{align}
    \hat{b}_{\omega}^R &= \int_0^{\infty} \D k \: \left( \alpha_{\omega k}^R \hat{a}_{k} + \beta_{\omega k}^R \hat{a}_{k}^\dd \right) ,
    \\
    \hat{b}_{\omega }^L &= \int_0^\infty \D k \: \left( \alpha_{\omega k}^L \hat{a}_{k} + \beta_{\omega k}^L \hat{a}_{k}^\dd \right) ,
\end{align}
where $\alpha_{\omega k} = \langle g_\omega , u_k \rangle$, $\beta_{\omega k} = \langle g_\omega^\star, u_k \rangle$. The Bogoliubov coefficients take the form \cite{crispino_2008RevModPhys.80.787} 
\begin{align}
    \alpha_{\omega k}^L &= \alpha_{\omega k}^{R\star } = \frac{i e^{\pi\omega/2a}}{2\pi \sqrt{\omega k}} \left( \frac{k}{a} \right)^{i\omega/a}\Gamma(1 - i \omega/a) ,
    \\
    \beta_{\omega k}^R &= \beta_{\omega k}^{L\star } = - \frac{i e^{-\pi\omega/2a}}{2\pi \sqrt{\omega k}} \left( \frac{k}{a} \right)^{i\omega/a} \Gamma(1 - i  \omega/a) ,
\end{align}
where $\Gamma(z)$ is the Gamma function \cite{gradshteyn2014table}. The existence of nonvanishing Bogoliubov coefficients between the
Rindler and Minkowski positive and negative frequency modes is a statement of the inequivalence of the Rindler and Minkowski vacua.
The Minkowski vacuum will contain zero Minkowski particles, but will contain Rindler particles. In particular, one finds that 
\begin{align}
    \langle 0_M | \hat{b}_\omega^{R\dd} \hat{b}_{\omega'}^R | 0_M \rangle &= \int\D k \: \beta_{\omega k}^{R\star} \beta_{\omega'k}^R ,
\end{align}
and after the substitution $z = \ln (k)$ one obtains the result 
\begin{align}
    \langle 0_M | \hat{b}_\omega^{R\dd} \hat{b}_{\omega'}^R | 0_M \rangle &= \frac{\delta(\omega - \omega')}{e^{2\pi\omega/a}-1} ,
    \label{eq24}
\end{align}
having used the identity $| \Gamma (1 + i x ) |^2 = \pi x \mathrm{csch}(\pi x)$. The formally divergent $\delta(\omega-\omega')$ appears because we have considered an infinite volume of space, and can be regularised by considering wavepacket modes. The important point is that one finds that the particle number for each Rindler frequency $\omega$ is given by a Planck distribution with temperature $T_U = a(2\pi)^{-1}$, known as the Unruh temperature. Finally, we remark that the signature of thermalisation, the Planck distribution in Eq.\ (\ref{eq24}), is obtained by computing the vacuum expectation value of the number operator for a fixed trajectory of the accelerated observer. However, nothing precludes us from using this quantity as a benchmark for thermalisation in cases where the observer's motion has quantum degrees of freedom (DoFs), i.e.\ superpositions of trajectories.

\subsection{Unruh Effect in Diamond Reference Frame}
In analogy to Rindler observers ``localised'' to the right or left Rindler wedge, the strict localisation of finite-lifetime observers to a spacetime diamond in Minkowski spacetime gives rise to the so-called diamond temperature. In Ref.\ \cite{su2016spacetime}, the physical interpretation of this temperature was given in terms of an energy-scaled detector whose rapidly changing proper time yielded a thermal response proportional to the diamond localisation scale. Here we briefly review the derivation of the diamond temperature via the Bogoliubov transformation method shown in Ref.\ \cite{su2016spacetime}.

In analogy to the Rindler case, it is convenient to work in (1+1)-dimensions. Without loss of generality,
let us consider the static case where $\xi = 0$, in which the line element is conformal to the
Minkowski metric. The field can be decomposed in the basis of diamond modes internal and external to the zeroth diamond:
\begin{align}
    & \hat{\Phi}(V) 
    \non 
    \\
    & \:\:\: = \int_0^\infty \D \omega \: \Big( g_\omega^{(0)} (V) \hat{b}_\omega^{(0)} + g_\omega^\mathrm{(ex)} (V) \hat{b}_\omega^\mathrm{(ex)}  + \mathrm{H.c} \Big) ,
\end{align}
where the diamond mode functions take the form \cite{ida2013modular}
\begin{align}
    g_\omega^{(0)} &= \frac{1}{\sqrt{4\pi\omega}} \left( \frac{1 + aV/2}{1 - aV/2} \right)^{-i\omega/a}
    ,
    \\
    g_\omega^\mathrm{(ex)} &= \frac{1}{\sqrt{4\pi\omega}} \left( \frac{aV/2 + 1}{aV/2 -1 } \right)^{i\omega/a} \Theta(|V| - 2/a) ,
\end{align}
where $\Theta(z)$ is the Heaviside step function, and we have considered the left-moving sector of the field only. The bosonic operators $\hat{b}_\omega^{(0)}$, $\hat{b}_\omega^\mathrm{(ex)}$ are single-frequency annihilation operators for the modes localised to the zeroth diamond and those external to it, respectively. These are directly analogous to the Rindler operators $\hat{b}_\omega^R$, $\hat{b}_\omega^L$.

The Bogoliubov coefficients between the diamond modes and the Minkowski modes can be obtained via the usual Klein-Gordon inner product \cite{su2016spacetime}, 
\begin{align}
    \alpha_{\omega k}^{(0)} &= \frac{1}{\pi a} \sqrt{\frac{\kappa}{\Omega}} \int_{-1}^{+1} \D s \: \left( \frac{1 + s}{1 - s} \right)^{i\Omega} e^{-2i\kappa s}
    \non 
    ,
    \\
    &= \frac{2}{a} \frac{\sqrt{\Omega\kappa}}{\sinh(\pi\Omega)} e^{2i\kappa} M(1 + i \Omega, 2 , -4 i \kappa ) , 
    \\
    \beta_{\omega k}^{(0)} &= - \frac{1}{\pi a} \sqrt{\frac{\kappa}{\Omega}} \int_{-1}^{+1} \D s \: \left( \frac{1 + s}{1 - s} \right)^{i\Omega}e^{2i\kappa s}
    \non 
    ,
    \\
    &= - \frac{2}{a} \frac{\sqrt{\Omega\kappa}}{\sinh(\pi\Omega)} e^{-2i\kappa} M(1 + i \Omega, 2, 4i\kappa ) , 
\end{align}
where $M(a,b,z)$ is the Kummer function (confluent hypergeometric function of the first kind) and $\Omega = \omega/a$, $\kappa = k/a$ \cite{gradshteyn2014table}. The particle number distribution in the diamond is given by 
\begin{align}
    \langle 0_M | \hat{b}_\omega^{(0)\dd } \hat{b}_{\omega'}^{(0)} | 0_M \rangle &= \int\D k \: \beta_{\omega k}^{(0)\star } \beta_{\omega' k}^{(0)} = \frac{\delta(\omega - \omega')}{e^{2\pi\omega/a}-1},
\end{align}
which again is a thermal distribution with temperature $T_D = a(2\pi)^{-1}$ in direct analogy with the Unruh temperature. A derivation of this result is shown in the Appendix. The interpretation of this result is that the tighter the localisation of the diamond, the more rapidly an observer's proper time will vary with respect to the global Minkowski time. In analogy with higher accelerations in the Rindler case, higher localisation leads to a higher observer temperature.

\subsection{Minkowski Vacuum as an Entangled State}
The derivations shown above reveal the property of the Minkowski vacuum state as an entangled state between the disjoint left and right Rindler wedges, or the interior and exterior of the spacetime diamond. It is well-known that the Minkowski vacuum is a two-mode squeezed state of the left and right Rindler modes (interior and exterior diamond modes) \cite{crispino_2008RevModPhys.80.787}. In the discrete-frequency approximation, the Minkowski vacuum can thus be expressed as follows:
\begin{align}
    | 0_M \rangle &= \prod_i C_i \sum_{n_i=0}^{+\infty} \frac{e^{-\pi n_i \omega_i/a}}{n_i!} \left( \hat{b}_{\omega_i}^{R\dd} \hat{b}_{\omega_i}^{L\dd} \right)^{n_i} | 0_R \rangle , 
    \non 
    \\
\label{eq45}
    &= \prod_i \left( C_i \sum_{n_i=0}^{+\infty} e^{-\pi n_i \omega_i/a} | n_i, R \rangle \otimes | n_i , L \rangle \right) 
\end{align}
(an analogous expression exists in terms of the interior and exterior diamond modes), where $C_i = \sqrt{1- e^{-2\pi\omega_i/a}}$ is a normalisation factor. In Eq.\ (\ref{eq45}), $|0_R \rangle$ is the Rindler vacuum state, while $|n_i,R\rangle$, $|n_i,L\rangle$ are $n_i$-particle states with frequency $\omega_i$ in the right and left wedges respectively. 
Equation \eqref{eq45} shows that the Minkowski vacuum state is an entangled state between the two Rindler wedges (interior--exterior of a diamond region).

An observer uniformly accelerating in the right wedge (confined to the zeroth diamond) cannot access the state in the  left wedge 
and so tracing out the state in the left wedge 
leaves a mixed state
\begin{align}\label{eq34}
    \hat{\rho}_R &= \prod_i \left( C_i^2 \sum_{n_i=0}^\infty e^{-\pi n_i \omega_i/a} | n_i, R \rangle\langle n_i , R | \right) 
\end{align}
(an analogous expression can be written in terms of interior and exterior diamond modes). This is of course a thermal state density matrix for a system of free bosons at the Unruh temperature, $T_U = a(2\pi)^{-1}$. 
We can thus refer to the entangled state \eqref{eq45} as a ``purification'' of the 
mixed state in \eqref{eq34}. We note that in the field of quantum information it is common to use the fact that for any mixed state one can formally define a (non-unique) purification--a pure state including ancillary degrees of freedom where the reduced state of one of the subsystems is the  original mixed state. This will be key for finding connection to the approach of Ref.~\cite{wood2021https://doi.org/10.48550/arxiv.2112.07860}. We further note that in the context of relativistic quantum field theory this structure naturally arises the other way around: while the field can be in the global vacuum state, an accelerating  observer interacts with modes confined e.g.~to the right Rindler wedge, where the reduced state is thermal as explained above. Introducing field modes in the left wedge purifies this state back into the Minkowski vacuum, which in this description is thus a two-mode squeezed state of Eq.\ (\ref{eq45}) \cite{LEE1986437,takagi10.1143/PTP.88.1}. 
Our interest in the remainder of this paper is in scenarios in which probes or observers interact with a superposition of states that are associated 
with different 
purifications, which we explain in detail below. 

\section{Superposition of Purifications in Rindler}
Let us now consider an observer travelling in a superposition of the trajectories shown in Fig.\ \ref{fig:rindlerspace}, each with the same proper acceleration but translated by a constant offset $s/a$ (according an inertial observer).

We introduce a quantum degree of freedom, $i$, that controls which of the trajectories the observer follows. Preparing the control in a superposition results in the observer following the two trajectories in superposition. We henceforth denote the two states of the control  by $| R\rangle$ and $| R' \rangle$ and consider that the initial state of the control and the field can be written as the product,
\begin{align}
    | \psi \rangle &= \frac{1}{\sqrt{2}} \left( | R \rangle + | R' \rangle \right) | 0_M \rangle .
\end{align}
We assume that the control states $| R \rangle , | R' \rangle$ are normalised and mutually orthogonal. Note that this does not necessitate that the field modes restricted to the Rindler wedges arising from these trajectories are orthogonal. We also define the following quantum-controlled annihilation operator,
\begin{align}
    \hat{b}_\omega &= \hat{b}_\omega^R \otimes | R \rangle \langle R | + \hat{b}_\omega^{R\prime} \otimes | R' \rangle \langle R' | .
\end{align}
Applying this to the initial state and conditioning on the control measured in a superposition state $| \eta \rangle = ( |R\rangle + |R ' \rangle )/\sqrt{2}$ (e.g.\ recombining the paths of an interferometer) gives the conditional state, 
\begin{align}
    \langle \eta | \hat{b}_\omega | \psi \rangle &= \frac{1}{2} ( \hat{b}_\omega^R + \hat{b}_\omega^{R\prime} ) \ranglem .
\end{align}
The particle number is thus a sum of four terms:
\begin{align}
    & \left| \langle \eta | \hat{b}_\omega | \psi \rangle \right|^2  
    \non 
    \\
    & \:\:\:  = \frac{1}{4} \Big[ \langle 0_M | \hat{b}_\omega^{R\dd} \hat{b}_{\omega'}^R |0_M \rangle 
    + \langle 0_M | \hat{b}_\omega^{R\dd} \hat{b}_{\omega'}^{R\prime} | 0_M \rangle 
    \non 
    \\
    & \qquad  + \langle 0_M | \hat{b}_\omega^{R\prime \dd} \hat{b}_{\omega'}^{R} | 0_M \rangle 
    + \langle 0_M | \hat{b}_\omega^{R\prime \dd} \hat{b}_{\omega'}^{R\prime} | 0_M \rangle \Big]  .\vphantom{\frac{1}{4}}
    \label{eq36}
\end{align}
The first and fourth terms are contributions to the particle number solely due to the local interaction with the field along the individual trajectories. Individually, these yield the usual Planckian spectrum derived previously. The second and third terms contain correlations between the shifted wedges, mixing the $\hat{b}_\omega^R$, $\hat{b}_\omega^{R\prime}$ operators. The general composition of Eq.\ (\ref{eq36})--two local terms describing contributions from the individual branches of the superposition, and two interference terms describing correlations between them--also appears in prior analyses of scenarios involving quantum-controlled thermalisation \cite{foo2020unruhdewitt,wood2021https://doi.org/10.48550/arxiv.2112.07860}. The interference terms are those that will allow us to make a direct connection between the two approaches. Let us now derive these interference terms explicitly.

We consider the shifted Rindler wedge as having the constant offset $s/a$ from the original one. The Bogoliubov transformation is given by 
\begin{align}
    \hat{b}_\omega^{R\prime} &= \int_0^\infty\D k \: \Big( \alpha_{\omega k}^{R\prime} \hat{a}_k + \beta_{\omega k}^{R\prime} \hat{a}_k^\dd \Big) 
    , 
    \\
    \hat{b}_\omega^{L\prime} &= \int_0^\infty\D k \: \Big( \alpha_{\omega k}^{L\prime} \hat{a}_k + \beta_{\omega k}^{L\prime} \hat{a}_k^\dd \Big) ,
\end{align}
where the Bogoliubov coefficients in the translated right Rindler wedge are related to those in the original wedge via
\begin{align} 
    \alpha_{k\omega}^{R\prime} &= e^{iks/a} \alpha_{k\omega}^{R} 
    \vt , 
    \label{eq37}
    \\
    \beta_{k\omega}^{R\prime} &= e^{iks/a} \beta_{k\omega}^{R} .
    \vt 
    \label{eq38}
\end{align}
The additional phases in Eq.\ (\ref{eq37}) and (\ref{eq38}) can be understood as breaking the symmetry of the shifted wedges with respect to the Minkowski coordinates with which the original Rindler wedges are defined. Owing to the translational invariance of a given wedge, the particle number in the shifted wedge also obeys a Planck distribution:
\begin{align}
    \langle 0_M | \hat{b}_\omega^{R\prime\dd} \hat{b}_{\omega'}^{R\prime} | 0_M \rangle = \frac{\delta(\omega - \omega')}{e^{2\pi\omega/a}-1} .
\end{align}
Now, for the observer in a superposition of wedges, the cross terms that mix $\hat{b}_\omega^R$, $\hat{b}_\omega^{R\prime}$ operators are given by $\langle 0_M| \hat{b}_\omega^{R\dd}\hat{b}_{\omega'}^{R\prime} | 0_M\rangle = \langle 0_M|\hat{b}_\omega^{R\prime\dd} \hat{b}_{\omega'}^R | 0_M\rangle$, 
\begin{align}
    \langle 0_M|\hat{b}_\omega^{R\dd} \hat{b}_{\omega'}^{R\prime} | 0_M\rangle 
    &= \int\D k \: \beta_{\omega k}^{R\star} \beta_{\omega'k}^{R\prime} 
    , 
\end{align}
which evaluate explicitly to 
\begin{align}
    & \langle 0_M | \hat{b}_\omega^{R\dd} \hat{b}_{\omega'}^{R\prime} | 0_M \rangle 
    \non \vt \\
    & \:\:\: = \Lambda_{\mathrm{RR}'} (s)^{i(\Omega - \Omega')} e^{\pi(\Omega- \Omega')/2} \Gamma\Big[ -i (\Omega - \Omega' ) \Big] ,
\end{align}
where
\begin{align}
     \Lambda_{\mathrm{RR}^\prime} = \frac{e^{-\pi(\Omega+\Omega')/2}}{4\pi^2  a\sqrt{\Omega\Omega'}} \Gamma(1+i\Omega) \Gamma(1-i\Omega') .
\end{align}
As discussed previously, this expression is formally divergent for $\Omega = \Omega'$, however this should be understood in a distributional sense: when a wavepacket mode is considered, the expression is regular. We find that the presence of these cross-terms, representing correlations between the two wedges, leads to a non-Planckian and nonthermal distribution. The implication is that an observer with a uniform acceleration, travelling on a trajectory in a superposition of spatial translations, will not see a thermalised vacuum state. This corroborates the result of \cite{foo2020unruhdewitt,fooPhysRevResearch.3.043056}, in which an Unruh-deWitt detector in such a configuration likewise did not exhibit a thermal response. Let us emphasise here that although prior works e.g.\ Ref.\ \cite{felcePhysRevLett.125.070603} report similar findings (Ref.\ \cite{felcePhysRevLett.125.070603} in the context of thermalising channels applied in a superposition of causal orders), the corroboration of this phenomenon in the present scenario (i) is not trivially expected, and (ii) sheds light on the physics underlying earlier results. Regarding (i), owing to the translational invariance of quantum field theory and in particular the vacuum state, one might expect that thermalisation does result, since the exact same channel is applied whether the control is in $| R \rangle$ or $| R ' \rangle$. Alternatively, one might suppose that from the reference frame of the detector, ``there is nothing in superposition'' except for the relative coordinates defining the origin, and likewise therefore, that thermalisation should result. For (ii), what our analysis demonstrates is that the inequivalent decompositions of the respective sets of Rindler modes $\hat b_\omega^R, \hat b_\omega^{R\prime}$ with respect to the global Minkowski modes $\hat a_k$ in essence ``fixes'' the origin of coordinates, through which the interference effects are defined with respect to. This is the case even though the coordinate origin is arbitrary for an observer on a classical trajectory. This latter point motivates interesting questions concerning the notion of classical thermodynamics in quantum reference frames, which we leave for future work. Finally, we remark that in the limit $s \to 0$ and $\omega \to \omega'$ (no superposition), one recovers the usual result and the particle number is Planckian. Likewise when $s \to \infty$, one can make a stationary phase approximation \cite{bleistein1975asymptotic} leading to the cross-terms vanishing.

\subsection{Nonorthogonal Rindler Vacua}
We will now discuss how one can intuitively understand the nonthermal particle number in terms of the different decompositions of the Minkowski vacuum shown in Eq.\ (\ref{eq45}), and make an explicit connection to the quantum-informatiom approach of Ref.~\cite{wood2021https://doi.org/10.48550/arxiv.2112.07860}) 
associated with different states of the control $| R \rangle, | R ' \rangle$,
\begin{align}
    | \psi \rangle &= \frac{1}{\sqrt{2}} \prod_i C_i \sum_{n_i=0}^{+\infty} e^{-\pi n_i \omega_i/a} | n_i, R \rangle \otimes | n_i , L \rangle \otimes | R \rangle 
    \non 
    \\
    & + \frac{1}{\sqrt{2}} \prod_i C_i \sum_{n_i=0}^{+\infty}  e^{-\pi n_i\omega_i/a} | n_i , R' \rangle \otimes | n_i , L' \rangle \otimes | R' \rangle .
\end{align}
Measuring the control in the superposition basis $| \eta \rangle = ( |R \rangle + |R ' \rangle)/\sqrt{2}$ and tracing out the DoFs associated with the left wedge(s), one finds the final state of the right wedge(s) to be, 
\begin{align}\label{eq47}
    & \mathrm{Tr}_\mathrm{L} \langle \eta |  \hat{\rho} | \eta \rangle 
    \non 
    \\
    & \:\:\: = \frac{1}{4} \prod_i C_i^2 \sum_{n_i=0}^{+\infty} e^{-2\pi n_i \omega_i/a} | n_i, R \rangle\langle n_i, R |
    \non 
    \\
    & \:\:\: + \frac{1}{4} \prod_{i,j}C_iC_j \sum_{n_i=0}^{+\infty} e^{-\pi(n_i\omega_i + n_j\omega_j)/a} | n_i , R \rangle\langle n_i,R' |  \Pi_{\mathrm{LL}\prime} 
    \non 
    \\
    & \:\:\: + \frac{1}{4} \prod_{i,j}C_iC_j \sum_{n_i=0}^{+\infty} e^{-\pi(n_i\omega_i + n_j\omega_j)/a} | n_i , R' \rangle\langle n_i,R |  \Pi_{\mathrm{LL}\prime} 
    \non 
    \\
    & \:\:\:  + \frac{1}{4} \prod_i C_i^2 \sum_{n_i=0}^{+\infty} e^{-2\pi n_i \omega_i/a} | n_i, R' \rangle\langle n_i, R'|,
\end{align}
where we have denoted 
\begin{align}\label{eq:PiLL}
    \Pi_{\mathrm{LL}\prime} &= \left| \langle n_i , L | n_j , L' \rangle \right|^2
\end{align} 
as the overlap of the translated left Rindler states. We see that the final state of the field contains two thermal contributions from the respective right wedges $R$, $R'$, as well as two cross terms due to interference between the modes in these wedges. When the wedges are infinitely separated ($s \to \infty$), the overlap between $| n_i, L \rangle$, $| n_j, L'\rangle$ vanishes, leaving a thermal state. This likewise corroborates the result obtained in \cite{foo2020unruhdewitt}.

In Ref.~\cite{wood2021https://doi.org/10.48550/arxiv.2112.07860}, two models for an operational understanding of superpositions of temperatures were discussed. The first is a probe thermalising with one of two systems (each, in general, in a different thermal state) depending on the state of a control degree of freedom; the second is a probe interacting with a bath whose state is a superposition of purifications corresponding to different temperatures. It is this second model that we can identify as underlying the concrete thermalisation channels arising when relativistic local probes interact with a quantum field, discussed in the present work.

Apart from the above correspondence of the general settings, the reduced state of the bath in the quantum information approach of Ref.\ \cite{wood2021https://doi.org/10.48550/arxiv.2112.07860} and the reduced state of the right Rindler wedge in Eq.~\eqref{eq47} also have direct correspondence. In particular we find the overlap between the field states from the two right wedges, Eq.~\eqref{eq:PiLL}, in the quantum information approach appearing as the overlap between the ancillary states belonging to the two considered purifications. Thermal weights appear likewise in the same manner in both approaches.

\section{Superposition of Diamond Purifications}
We can perform an analogous calculation for the case of an observer in a superposition of localised diamond trajectories. We consider an observer in a quantum superposition of static trajectories in the zeroth and $n$th diamond (illustrated in Fig.~\ref{fig:diamond illustration}), 
\begin{align}
    |\psi \rangle &= \frac{1}{\sqrt{2}} \left( | 0\rangle + | n \rangle \right) | 0_M \rangle. 
\end{align}
The observer interacts with the mode 
\begin{align}
    \hat{b}_\omega &= \hat{b}_\omega^{(0)} \otimes | 0\rangle\langle 0| + \hat{b}_\omega^{(n)} \otimes | n \rangle \langle n | ,
\end{align}
giving a detected particle number of the same form as Eq.\ (\ref{eq36}). Now, the shifted diamond modes are related to the Minkowski modes via the Bogoliubov transformations, 
\begin{align}
    \hat{b}_\omega^{(n)} &= \int_0^\infty\D k \: \Big( \alpha_{\omega k}^{(n)} \hat{a}_k + \beta_{\omega k}^{(n)} \hat{a}_k^\dd \Big) , 
    \\
    \hat{b}_\omega^{(n,\mathrm{ex})} &= \int_0^\infty \D k \: \Big( \alpha_{\omega k}^{(n,\mathrm{ex})} \hat{a}_k + \beta_{\omega k}^{(n,\mathrm{ex})} \hat{a}_k^\dd \Big) . 
\end{align}
Like the Rindler case, the Bogoliubov coefficients are related to those of the zeroth diamond via, 
\begin{align}
    \alpha_{\omega k}^{(n)} &= e^{-4i\kappa n} \alpha_{\omega k}^{(0)} 
    , 
    \\
    \beta_{\omega k}^{(n)} &= e^{4i\kappa n} \beta_{\omega k}^{(0)} .
\end{align}
Again, the translational invariance of the diamonds results in a thermal state inside the $n$th diamond at the diamond temperature $T_D = a(2\pi)^{-1}$:
\begin{align}
    \langle 0_M | \hat{b}_\omega^{(n)\dd}  \hat{b}_{\omega'}^{(n)} | 0_M \rangle &= \frac{\delta(\omega - \omega')}{e^{2\pi\omega/a}-1}.
\end{align}
For a superposition of spatially translated diamond modes, the cross terms mixing the $\hat{b}_\omega^{(0)}$, $\hat{b}_\omega^{(n)}$ operators take the form, 
\begin{align}
    \langle 0_M | \hat{b}_\omega^{(0)\dd } \hat{b}_{\omega'}^{(n)} | 0_M \rangle
    &= \int\D k \: \beta_{\omega k}^{(0)\star } \beta_{\omega'k}^{(n)}, 
\end{align}
which evaluates explicitly to
\begin{align} 
    \label{eq59}
    &= \Lambda_{\mathrm{nn}^\prime}  \left( \alpha - 1 \right)^{i \Omega - 1} \left( \alpha  + 1 \right)^{-1 - i \Omega' }  
     \non 
     \\
     & \:\:\: \times F \left[ 1 - i \Omega, 1 + i \Omega';2; - \left( \alpha -1 \right)^{-1} \left( -\alpha +1  \right)^{-1} \right] 
\end{align}
where $\alpha = -(2i/a)(1 + 2n)$, $F(\alpha,\beta;\chi,z)$ is the hypergeometric function \cite{gradshteyn2014table}, and we have defined,
\begin{align}
    \Lambda_{\mathrm{nn}^\prime} &= \frac{4}{a^3}\frac{\sqrt{\Omega\Omega'}\left( \alpha \right)^{-i(\Omega - \Omega')} }{\sinh(\pi\Omega)\sinh(\pi\Omega')} .
\end{align}
Again, the complicated correlations between diamonds leads to cross terms that are characteristically nonthermal. This is another manifestation of the effect demonstrated in the Rindler case--even though individually the diamonds give rise to a thermal state at the same temperature, their superposition leads to a lack of thermalisation. The expression in Eq.~(\ref{eq59}) vanishes in the limit $n \to \infty$, while it can be shown, using the integral version of the Bogoliubov coefficients, that it reduces to the single-diamond contribution in the limit $n \to 0$, when the diamonds overlap. Before concluding this section, we remark that we expect similar results to hold for superposed diamond observers in (3+1)-dimensions. This is because the results derived here solely rely on the mode structure between different spacetime regions and their respective overlaps. We expect this general structure to carry over to higher dimensions.

\section{Superpositions of Orthogonally Translated Modes}
We have demonstrated that observers interacting with thermal reduced state of a field nevertheless do not in general witness a thermal particle distribution if the reduced state is associated with different purifications in superposition. We traced the lack of thermality to the fact that the field modes associated with the supersposed purifications are neither equal nor orthogonal. To highlight this result, we now consider some special cases in which a probe is in a superposition of trajectories, thus interacting with a state in a superposition of purifications, does thermalise. 

\subsection{Antiparallel Purifications in Rindler}
By a similar calculation as shown for the case of superposed Rindler trajectories, in general an observer accelerating in a superposition of ``antiparallel'' accelerations (i.e.\ in opposite directions) with some constant offset $s$ in the null coordinate will in general not thermalise. 
However a special case occurs when the offset is such that the Rindler wedges associated with the trajectories do not overlap but share a common origin. 
In such a case, the cross terms in the particle number distribution are given by 
\begin{align}
    \int\D k \: \beta_{\omega k}^{R\star} \beta_{\omega' k}^{L} &= \Lambda_\mathrm{RL} \delta(\Omega +\Omega') = 0,
    \vt 
\end{align}
where
\begin{align}
    & \Lambda_\mathrm{RL} 
    \non 
    \\
    & \:\:\: = \frac{e^{-\pi(\Omega+\Omega')/2}}{4\pi^2a \sqrt{\Omega\Omega'}} \Gamma(1+i\Omega) \Gamma(1 + i \Omega') (a)^{i(\Omega+\Omega')} .
\end{align}
In this special case, the cross terms vanish. This can equally be understood in terms of the vanishing Klein-Gordon inner product, $\langle g_\omega^R, g_\omega^L \rangle = 0$, between the mode functions in the right and left Rindler wedges. That is, since the wedges define fully disjoint Hilbert spaces, the modes are orthogonal to each other. The implication of this is that the total particle number in the Minkowski vacuum reduces to 
\begin{align}
    \langle 0_M | \hat{b}_\omega^\dd \hat{b}_{\omega'} | 0_M \rangle &= \frac{1}{2} \frac{\delta(\omega - \omega')}{e^{2\pi\omega/a}-1},
\end{align}
which is half of that obtained for an observer on a single classical trajectory in the right or left wedge. As before, we can understand this result in terms of the decomposition of the Minkowski vacuum into the purification states $| n_i, R \rangle$, $|n_i, L\rangle$. In this case, the accelerated observer interacts with the field in a quantum-controlled superposition of two identical purifications $|n_i,R\rangle \otimes | n_i , L \rangle$ 
\begin{align}
    | \psi \rangle &=  \prod_{i} C_i \sum_{n_i=0}^{+\infty} e^{-\pi n_i\omega_i/a} |n_i, R \rangle \otimes | n_i , L \rangle \otimes \frac{ | R \rangle + | L \rangle  }{\sqrt{2}} .
\end{align}
Depending on the state of the control, the observer then interacts with modes in the right or the left wedge. However since they are identical, the reduced state of the probe is independent of the wedge, and the cross terms are of the same form as the diagonal ones.
%
%
As the diagonal terms are thermal at a common temperature determined by the acceleration $a$, the whole state is likewise thermal.

\subsection{Orthogonally Translated Rindler Trajectories}
The previous example showed that for purification states that are orthogonal, the superposition of those states gives a thermal state. Quantum probes that interact with such superposition states will thermalise, in contrast to other examples in which the purifications possess some nontrivial overlap. 

\begin{figure}[h]
    \centering
    \includegraphics[width=0.7\linewidth]{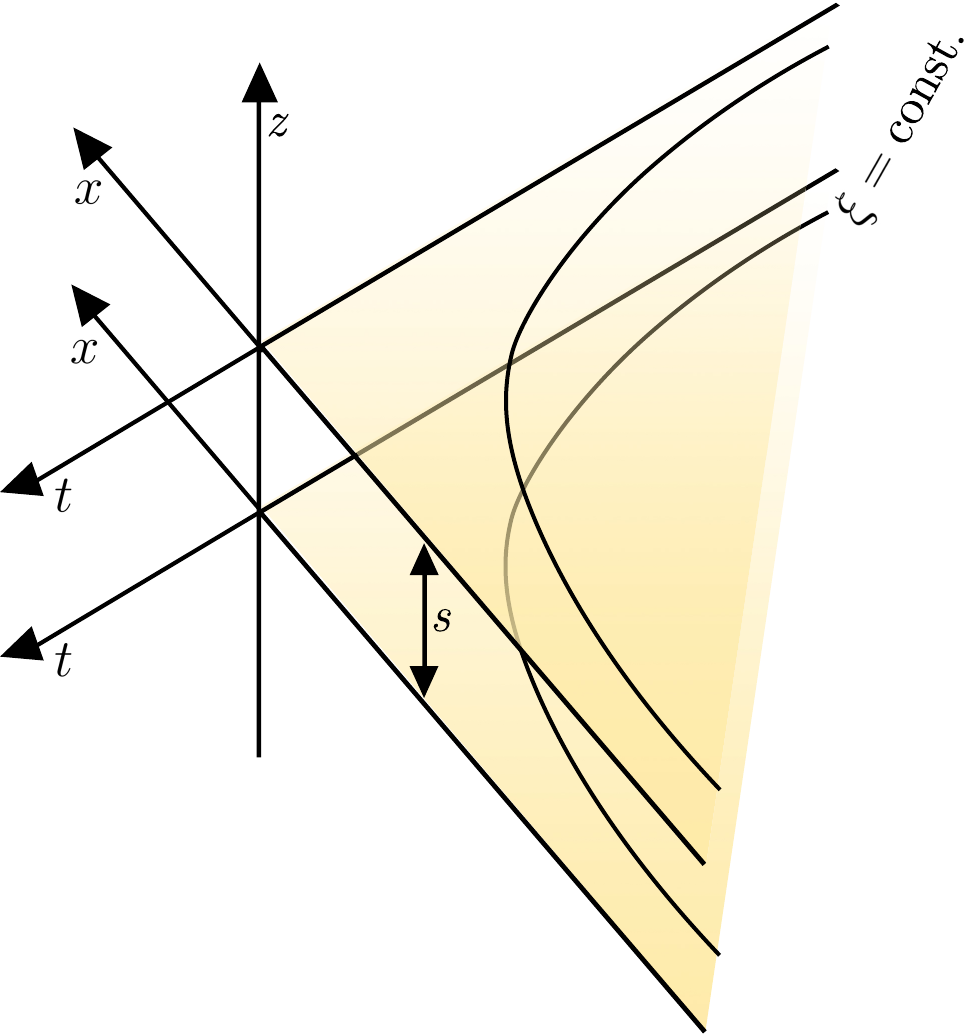}
    \caption{Schematic diagram of the orthogonally translated Rindler wedges.}
    \label{fig:orthogonal}
\end{figure}

Next, we can consider a superposition of two Rindler trajectories that are spatially separated in the direction orthogonal to motion. By the intuition developed thus far, we expect here that the particle distribution detected by observers travelling on such quantum trajectories will be thermal at the Unruh temperature associated with the Rindler wedge(s). Consider trajectories parametrised by the coordinates,
\begin{align}
    t_1 &= t_2 = a^{-1} e^{a\xi} \sinh(a\tau) , 
    \vt 
    \\
    x_1 &= x_2 = a^{-1} e^{a\xi} \cosh(a\tau) ,
    \vt 
\end{align}
while the coordinates in the orthogonal directions $(y,z)$ are offset by the constants $s$: 
\begin{align}
    y_1 &= y_2 + s = \mathrm{constant}
    , 
    \vt 
    \\
    z_1 &= z_2 + s = \mathrm{constant.}
    \vt 
\end{align}
The physical configuration of the respective wedges is illustrated in Fig.\ \ref{fig:orthogonal}.
In both cases, the (3+1)-dimensional scalar field can be expanded in the Rindler modes, 
\begin{align}
    \hat{\Phi} &= \int_0^\infty \D \omega \int\D^2 \textbf{k}_\perp \: \Big( v_{\omega \textbf{k}_\perp}^R (\tau, \xi, \textbf{x}_{\perp i}) \hat{a}_{\omega \textbf{k}_{\perp i}}^R 
    \non 
    \\
    & \qquad + v_{\omega \textbf{k}_\perp}^L (\bar{\tau}, \bar{\xi}, \textbf{x}_{\perp i}) \hat{a}_{\omega \textbf{k}_{\perp i}}^L + \mathrm{H.c} \Big) 
\end{align}
where $\textbf{x}_{\perp i} = (y_i,z_i)$ and $\textbf{k}_\perp = (k_{y_i}, k_{z_i})$. The mode functions on the future Killing horizon, $t = z$, $t>0$, are given by \cite{crispino_2008RevModPhys.80.787}
\begin{align}
    v_{\omega \textbf{k}_\perp}^R &= - \frac{i e^{i\textbf{k}_\perp \cdot \textbf{x}_\perp}}{4\pi \sqrt{a \sinh(\pi\omega/a)}} \frac{(\kappa/2a)^{-i\omega/a} e^{-i\omega v}}{\Gamma(1-i\omega/a)}
\end{align}
where $\kappa = \sqrt{k_{y_i}^2 + k_{z_i}^2}$. Using this form of the mode function, one can obtain the Bogoliubov coefficients: 
\begin{align}
\label{eq68}
    \alpha_{\omega k_x k_\perp}^R &= \frac{e^{\pi\omega/2a}}{\sqrt{4\pi k_0 a \sinh(\pi \omega/a)}} \left( \frac{k_0 + k_x}{k_0 - k_x} \right)^{-i\omega/2a} , 
    \\ 
    \label{eq69}
    \beta_{\omega k_x k_\perp }^R &= - \frac{e^{-\pi\omega/2a}}{\sqrt{4\pi k_0a \sinh(\pi\omega/a)}} \left( \frac{k_0 + k_x}{k_0 - k_x} \right)^{-i\omega/2a} ,
\end{align}
where $k_0 = \sqrt{k_x^2 + k_\perp^2}$. The Bogoliubov coefficients in the right wedge are related to those in the left by $\alpha_{\omega k_x k_\perp}^R = \alpha_{\omega - k_x k_\perp}^L$ and $\beta_{\omega k_x k_\perp}^R =  \beta_{\omega - k_x k_\perp}^L$. This further implies the relationship, 
\begin{align}\label{eq70}
    \left( \hat{a}_{\omega \textbf{k}_\perp}^R - e^{-\pi\omega/a} \hat{a}_{\omega - \textbf{k}_\perp}^{L\dd} \right) | 0_M \rangle &= 0 , 
    \vt 
    \\
    \label{eq71}
    \left( \hat{a}_{\omega \textbf{k}_\perp}^L - e^{-\pi\omega/a} \hat{a}_{\omega - \textbf{k}_\perp}^{R\dd } \right)  |0_M \rangle  &= 0 . 
\end{align}
Equations Eq.\ (\ref{eq70}) and (\ref{eq71}) are sufficient for demonstrating the Unruh effect \cite{crispino_2008RevModPhys.80.787}.\footnote{The analogous relationship in (1+1)-dimensions is
\begin{align}
    \left( \hat{b}_\omega^R - e^{-\pi\omega/a} \hat{a}_\omega^{L\dd} \right) | 0_M \rangle &= 0 ,
    \nonumber 
    \\
    \left( \hat{b}_\omega^L - e^{-\pi\omega/a} \hat{b}_\omega^{R\dd} \right) | 0_M \rangle &= 0 .
    \nonumber 
\end{align}} 
What is more important to note here is that Eq.\ (\ref{eq68}) and (\ref{eq69}) are invariant under translations in $(y,z)$--such a translation only introduces a global phase that has no effect on the final result. This means that an observer interacting with a superposition of modes localised to orthogonally translated wedges will see a thermal distribution of particles in the Minkowski vacuum. Though there may be complicated interference effects between the modes, such interference does not affect the dynamics in the plane of motion, which is the only relevant one for computing thermalisation associated with the Unruh effect. This is complementary to the previous result showing lack of thermalisation for Rindler wedges translated in the plane of motion. In that case, the translation of wedges with respect to the origin of Minkowski coordinates breaks the symmetry of the respective field modes, leading to a nontrivial overlap between them (i.e.\ a nonvanishing Klein-Gordon inner product).  

In contrast, a translation in the plane of motion (i.e.\ by some constant offset $s$ in the $x$-direction) gives Bogoliubov coefficients \cite{crispino_2008RevModPhys.80.787}
\begin{align}
    \alpha_{\omega k_x k_\perp} &= e^{-i(k_0 - k_z)s/2a} \alpha_{\omega k_x k_\perp} 
    , 
    \vt 
    \\
    \beta_{\omega k_x k_\perp} &= e^{+i(k_0-k_z)s/2a} \beta_{\omega k_x k_\perp} .
    \vt 
\end{align}
These relations are analogous to those derived in the (1+1)-dimensional case. Thus for a superposition of these modes with those in the original Rindler wedge, Eq.\ (\ref{eq70}) and (\ref{eq71}) will in general not be satisfied, and the particle number will be nonthermal.

\subsection{Unruh-deWitt Detector in de Sitter Spacetime}
Let us conclude this section with two applications of our results. We draw upon insights gained from the well-known Unruh-deWitt model, which in its most idealised form considers a pointlike two-level system linearly coupled to the field, through which it can experience transitions from its ground to excited state \cite{birrell1984quantum,pozasPhysRevD.94.064074,Louko_2008,stritzelbergerPhysRevD.101.036007,hendersonPhysRevLett.125.131602}. Recently, we extended this model to include quantum degrees of freedom in the detector's motion, allowing it to be prepared in a coherent superposition of different trajectories \cite{foo2020unruhdewitt, fooPhysRevResearch.3.043056}. 

The first example of a superposed detector that thermalises to the temperature of radiation in its environment is one that is situated in an expanding de Sitter universe (parametrised by static coordinates) and travelling in a superposition of worldlines rotated by some angle $\varphi$, such that the constant Euclidean distance between the two worldlines is $s = 2R \sin(\varphi/2)$ \cite{tian2016detecting,HUANG2017458}. Such a scenario is very similar to the previously considered example of a Rindler observer in a superposition of orthogonal translations (via the conformal equivalence of the Rindler and static de Sitter spacetimes \cite{griffiths2009exact}). Since the radial waveform of the field modes is unaffected by such a rotation ($\D \varphi$ being an orthogonal unit vector to $\D R$ in the static de Sitter metric), the detector will effectively interact with an orthogonal set of modes in superposition. 

Now, a detector on a classical worldline in this spacetime will exhibit a thermal response at the Gibbons-Hawking temperature, $T_G = \kappa(2\pi)^{-1}$, where $\kappa = \sqrt{l^{-2} - R^2}$ is the local surface gravity at the radial distance $R$ from the cosmological horizon, with $l$ being the characteristic length of the spacetime. This temperature is a combination of local acceleration effects and radiation effects due to the cosmological horizon. For a detector in the above mentioned superposition of angular coordinates $\varphi$, its response is given by \cite{Foo_2021schrodingers}
\begin{align}
    & F(\Omega) 
    \non 
    \\
    & \:\:\: = \frac{\Omega}{4\pi} \frac{1}{e^{2\pi\Omega/\kappa}-1} \left[ 1 + \frac{\sin \left(\frac{2\Omega}{\kappa} \mathrm{arcsinh}(s\kappa/2) \right)}{s \Omega \sqrt{1 + (s \kappa/2)^2}} \right] , \label{eq76}
\end{align}
where $\Omega$ is the energy gap of the detector. The response is a sum of ``local'' contributions from the individual branches of the superposition (the first term in the brackets) and an interference term (the second term) between the two paths. Thermalisation of a quantum probe with the temperature of its environment is characterised by the detailed balance form of the Kubo-Martin-Schwinger (KMS) condition, $F(\Omega) /F(-\Omega) = \exp ( - 2\pi\Omega/\kappa)$ \cite{martinPhysRev.115.1342}. The response of the superposed detector satisfies this condition, indicating that while its spectrum is not the usual Planckian one, it still thermalises to a well-defined temperature. We finally note that Eq.\ (\ref{eq76}) is identical (upon replacing the surface gravity $\kappa$ with the proper acceleration) to the response of a detector that is uniformly accelerating in a superposition of orthogonally translated Rindler wedges, considered previously.   

Before considering a final example, we point out an equivalent signature of thermal states in QFT is that the two-point function of the field is periodic in imaginary time, when pulled back to the fixed trajectory of the probe. For a fixed trajectory in the static de Sitter coordinates considered in this section, this two-point function takes the form, 
\begin{align}
    W ( \tau , \tau' ) &= - \frac{1}{16\pi^2} \frac{\kappa^2}{\sinh^2( \kappa ( \tau - \tau' - i \varepsilon) /2 )}
\end{align}
where $\tau, \tau'$ are proper times, and $\varepsilon$ is a UV-regulator, which clearly satisfies $W( \tau , \tau' ) \equiv W( \Delta \tau )$, $W( \Delta \tau - i \beta ) = W ( - \Delta \tau)$ where $\beta = 2\pi/\kappa$. However in the case of a probe in superposition of trajectories, there is not a single Wightman function that charaterises its response, but rather a ``controlled superposition'' (i.e.\ a coherent sum) of Wightman functions--two evaluated locally along the respective trajectories in superposition, and two evaluated nonlocally, giving correlations between the trajectories--whose Fourier transform gives Eq.\ (\ref{eq76}) (see also Ref.\ \cite{Foo_2021schrodingers} and Appendix). Notably, the ``nonlocal'' Wightman functions, 
\begin{align}
    & W_\mathrm{nl} (\tau, \tau') 
    \nonumber \vt \\
    &= - \frac{(\kappa^2/16\pi)}{\sinh^2(\kappa ( \tau - \tau' - i\varepsilon)/2 ) - \kappa^2 R^2 \sin^2 (\varphi/2)}
\end{align}
are also periodic in imaginary time. While we have demonstrated that the resulting detector response, Eq.\ (\ref{eq76}), satisfies the detailed balance condition, we contend that this result is not trivially expected, due to the fact that the response is sum of these local and nonlocal ``interference'' terms.

\subsection{Unruh-deWitt Detector in the BTZ Spacetime}
In a similar way, a detector in a superposition of angular separations, $\delta\phi$, around a (2+1)-dimensional Banados-Teitelboim-Zanelli (BTZ) black hole also thermalises to a well-defined temperature when interacting with a scalar field. On a fixed classical worldline outside such a black hole with mass $M$, the detector's response takes the form \cite{henderson2019btz,lifschitzPhysRevD.49.1929,henderson2018harvesting}
\begin{align}\label{eq86}
    F(\Omega) &= \frac{\sqrt{\pi}}{2} \frac{1}{e^{\Omega/T}+1} \sum_{n=-\infty}^{n=+\infty} P_{-\frac{1}{2} + \frac{i\Omega}{2\pi T}} ( \cosh \alpha_n^- ) ,
\end{align}
where $P_\nu(\cosh\alpha)$ is the associated Legendre function, and we have defined 
\begin{align}
    \cosh\alpha_n^- &= (1 + X) \cosh(2\pi n \sqrt{M} ) - X,
\end{align}
and $X = 4\pi^2l^2T^2$ where $l$ is the AdS length, and $T = \sqrt{M}l /(2\pi l^2 \sqrt{M^2/l^2-R}$ is the local temperature at the radial distance $R$ from the black hole horizon $r_H = \sqrt{M}l$. Despite its unusual form, Eq.\ (\ref{eq86}) satisfies the KMS detailed balanced condition--the detector thermalises to the temperature of its environment. For a detector in a superposition of angular coordinates around the black hole, its response takes the form
\begin{align}
        F(\Omega) &= \frac{\sqrt{\pi}}{4} \frac{1}{e^{2\Omega/T} + 1}  \sum_{n=-\infty}^{n = + \infty} \Big( P_{- \frac{1}{2} + \frac{i\Omega}{2\pi T}} ( \cosh \alpha_n^- ) 
        \non 
        \\
        & \qquad +  P_{-1/2 + \frac{i\Omega}{2\pi T}} ( \cosh \beta_n^- ) \Big) ,
\end{align}
where we have defined
\begin{align}
    \cosh\beta_n^- &= (1 + X) \cosh(\sqrt{M} (\delta \phi - 2\pi n ) ) - X .
    \vt 
\end{align}
Using the property $P_{-1/2+i\lambda}(\cosh\alpha) = P_{-1/2 - i \lambda}(\cosh\alpha)$, then we find that the detector response satisfies the detailed balance condition, $F(\Omega ) / F(-\Omega) = \exp ( - \Omega/T )$, indicating a thermal response to the Hawking radiation at temperature $T$. While the angular separation between the superposed trajectories introduces nontrivial interference between the modes, these do not perturb the detector response away from thermality. This is because the modes parametrised by different values of $\phi$ are orthogonal, due to the axial symmetry of the spacetime \cite{lifschitzPhysRevD.49.1929}. This is analogous to the de Sitter case previously considered. 

\section{Conclusion}
In this paper, we have pinpointed the physical origin of recent results 
where quantum probes interacting with a superposition of seemingly invariant states of environment nevertheless do not thermalise.  
We have shown that the fundamental reason for this nonthermalisation is the fact that for each amplitude the probe interacts with a different set of modes, eg modes localised in a Rindler wedge arising from a given accelerated motion. For spatial superpositions of such motions, the relevant field modes with which the probe can interact are incommensurate and correlated in a complicated manner, resulting in general in a different than Planckian distribution of particles accessible to such a superposed observer or probe. As a result,  dynamics of any probe interacting with a field in a such a scenario would lack a detailed balanced-satisfying response, as shown in the present work on concrete examples. On the other hand, when the involved modes are orthogonal or identical, thermality is recovered.

The relevance of the different sets of modes defining the accessible  environment to the relativistic probe in superposition is also what connects relativistic scenarios with the quantum-information theoretic approach to the notion of temperature and superpositions of thermalisations from Ref.~\cite{wood2021https://doi.org/10.48550/arxiv.2112.07860}. Indeed, one of the proposed models for an operational scenario where a superposition of thermalisations could arise was defined via a probe interacting with a bath which, when purified, was represented as a superposition of different pure states (different purifications). The scenarios discussed in the present work reveal that this seemingly artificial construction can in fact be natural or even ubiquitous in relativistic physics.We emphasise that by utilising a concrete setup of a probe interacting with a relativistic quantum field (and thereby its underlying mode structure), we have explicitly highlighted the mechanism for the probe's thermalisation or nonthermalisation. This is in contrast with prior works employing generic Kraus operator decompositions for the thermal channel (e.g.\ \cite{felcePhysRevLett.125.070603,wood2021https://doi.org/10.48550/arxiv.2112.07860}), where the quantum-mechanical mode structure of the bath is not explicitly considered. Therefore the application of those frameworks to the specific scenarios considered here (i.e.\ involving observers following particular trajectories in relativistic spacetime) is limited.


The framework we have utilised here uniquely combines aspects of quantum thermodynamics, field theory, and relativity. We anticipate that here developed tools and ideas will find application in developing further understanding of thermodynamics and interactions more generally, in quantum reference frames, and in scenarios involving indefinite metrics \cite{giacomini2019,paczos2022quantum,debski2022universality,delahamette2021falling,giacomini10.1116/5.0070018}. Our results also motivate additional unexplored questions at the intersection of these reseach fields. For example, it is not currently understood how a quantum-delocalised system would respond to a thermal environment whose temperature locally varies. Such a scenario presents an interesting topic for future investigation. 

\section{Acknowledgements}

\noindent This research was supported by the Australian Research Council Centre of Excellence for Quantum Computation and Communication Technology (Project No. CE170100012).

\bibliographystyle{quantum}
\bibliography{main}

\begin{thebibliography}{10}

\bibitem{kosloffe15062100}
Ronnie Kosloff.
\newblock ``Quantum thermodynamics: A dynamical viewpoint''.
\newblock \href{https://dx.doi.org/10.3390/e15062100}{Entropy {\bf 15}, 2100--2128}~(2013).

\bibitem{andersdoi:10.1080/00107514.2016.1201896}
Sai Vinjanampathy and Janet Anders.
\newblock ``Quantum thermodynamics''.
\newblock \href{https://dx.doi.org/10.1080/00107514.2016.1201896}{Contemporary Physics {\bf 57}, 545--579}~(2016).

\bibitem{Lindblad1976}
G.~Lindblad.
\newblock ``On the generators of quantum dynamical semigroups''.
\newblock \href{https://dx.doi.org/10.1007/BF01608499}{Communications in Mathematical Physics {\bf 48}, 119--130}~(1976).

\bibitem{gorinidoi:10.1063/1.522979}
Vittorio Gorini, Andrzej Kossakowski, and E.~C.~G. Sudarshan.
\newblock ``Completely positive dynamical semigroups of n‐level systems''.
\newblock \href{https://dx.doi.org/10.1063/1.522979}{Journal of Mathematical Physics {\bf 17}, 821--825}~(1976).
\newblock  \href{http://arxiv.org/abs/https://pubs.aip.org/aip/jmp/article-pdf/17/5/821/19090720/821\_1\_online.pdf}{arXiv:https://pubs.aip.org/aip/jmp/article-pdf/17/5/821/19090720/821\_1\_online.pdf}.

\bibitem{breuer2002theory}
Heinz-Peter Breuer, Francesco Petruccione, et~al.
\newblock ``The theory of open quantum systems''.
\newblock \href{https://dx.doi.org/https://doi.org/10.1093/acprof:oso/9780199213900.001.0001}{Oxford University Press on Demand}. ~(2002).

\bibitem{gardiner2015quantum}
Crispin Gardiner and Peter Zoller.
\newblock ``The quantum world of ultra-cold atoms and light book ii: the physics of quantum-optical devices''.
\newblock \href{https://dx.doi.org/https://doi.org/10.1142/p983}{Volume~4}.
\newblock World Scientific Publishing Company. ~(2015).

\bibitem{calderbankPhysRevA.54.1098}
A.~R. Calderbank and Peter~W. Shor.
\newblock ``Good quantum error-correcting codes exist''.
\newblock \href{https://dx.doi.org/10.1103/PhysRevA.54.1098}{Phys. Rev. A {\bf 54}, 1098--1105}~(1996).

\bibitem{howl2022quantum}
Richard Howl, Ali Akil, Hlér Kristjánsson, Xiaobin Zhao, and Giulio Chiribella.
\newblock ``Quantum gravity as a communication resource''~(2022).
\newblock  \href{http://arxiv.org/abs/2203.05861}{arXiv:2203.05861}.

\bibitem{procopioPhysRevA.101.012346}
Lorenzo~M. Procopio, Francisco Delgado, Marco Enr\'{\i}quez, Nadia Belabas, and Juan~Ariel Levenson.
\newblock ``Sending classical information via three noisy channels in superposition of causal orders''.
\newblock \href{https://dx.doi.org/10.1103/PhysRevA.101.012346}{Phys. Rev. A {\bf 101}, 012346}~(2020).

\bibitem{Abbott2020communication}
Alastair~A. Abbott, Julian Wechs, Dominic Horsman, Mehdi Mhalla, and Cyril Branciard.
\newblock ``Communication through coherent control of quantum channels''.
\newblock \href{https://dx.doi.org/10.22331/q-2020-09-24-333}{{Quantum} {\bf 4}, 333}~(2020).

\bibitem{gaoPhysRevLett.124.030502}
Yu~Guo, Xiao-Min Hu, Zhi-Bo Hou, Huan Cao, Jin-Ming Cui, Bi-Heng Liu, Yun-Feng Huang, Chuan-Feng Li, Guang-Can Guo, and Giulio Chiribella.
\newblock ``Experimental transmission of quantum information using a superposition of causal orders''.
\newblock \href{https://dx.doi.org/10.1103/PhysRevLett.124.030502}{Phys. Rev. Lett. {\bf 124}, 030502}~(2020).

\bibitem{Paunkovic2020causalorders}
Nikola Paunkovi{\'{c}} and Marko Vojinovi{\'{c}}.
\newblock ``Causal orders, quantum circuits and spacetime: distinguishing between definite and superposed causal orders''.
\newblock \href{https://dx.doi.org/10.22331/q-2020-05-28-275}{{Quantum} {\bf 4}, 275}~(2020).

\bibitem{dahlstenPhysRevLett.129.230604}
Xiangjing Liu, Daniel Ebler, and Oscar Dahlsten.
\newblock ``Thermodynamics of quantum switch information capacity activation''.
\newblock \href{https://dx.doi.org/10.1103/PhysRevLett.129.230604}{Phys. Rev. Lett. {\bf 129}, 230604}~(2022).

\bibitem{BAN2021127381}
Masashi Ban.
\newblock ``Non-classicality created by quantum channels with indefinite causal order''.
\newblock \href{https://dx.doi.org/https://doi.org/10.1016/j.physleta.2021.127381}{Physics Letters A {\bf 402}, 127381}~(2021).

\bibitem{felcePhysRevLett.125.070603}
David Felce and Vlatko Vedral.
\newblock ``Quantum refrigeration with indefinite causal order''.
\newblock \href{https://dx.doi.org/10.1103/PhysRevLett.125.070603}{Phys. Rev. Lett. {\bf 125}, 070603}~(2020).

\bibitem{niePhysRevLett.129.100603}
Xinfang Nie, Xuanran Zhu, Keyi Huang, Kai Tang, Xinyue Long, Zidong Lin, Yu~Tian, Chudan Qiu, Cheng Xi, Xiaodong Yang, Jun Li, Ying Dong, Tao Xin, and Dawei Lu.
\newblock ``Experimental realization of a quantum refrigerator driven by indefinite causal orders''.
\newblock \href{https://dx.doi.org/10.1103/PhysRevLett.129.100603}{Phys. Rev. Lett. {\bf 129}, 100603}~(2022).

\bibitem{Chiribella_2021}
Giulio Chiribella, Manik Banik, Some~Sankar Bhattacharya, Tamal Guha, Mir Alimuddin, Arup Roy, Sutapa Saha, Sristy Agrawal, and Guruprasad Kar.
\newblock ``Indefinite causal order enables perfect quantum communication with zero capacity channels''.
\newblock \href{https://dx.doi.org/10.1088/1367-2630/abe7a0}{New Journal of Physics {\bf 23}, 033039}~(2021).

\bibitem{CHAPEAUBLONDEAU2022128300}
François Chapeau-Blondeau.
\newblock ``Indefinite causal order for quantum metrology with quantum thermal noise''.
\newblock \href{https://dx.doi.org/https://doi.org/10.1016/j.physleta.2022.128300}{Physics Letters A {\bf 447}, 128300}~(2022).

\bibitem{guhaPhysRevA.102.032215}
Tamal Guha, Mir Alimuddin, and Preeti Parashar.
\newblock ``Thermodynamic advancement in the causally inseparable occurrence of thermal maps''.
\newblock \href{https://dx.doi.org/10.1103/PhysRevA.102.032215}{Phys. Rev. A {\bf 102}, 032215}~(2020).

\bibitem{simonovPhysRevA.105.032217}
Kyrylo Simonov, Gianluca Francica, Giacomo Guarnieri, and Mauro Paternostro.
\newblock ``Work extraction from coherently activated maps via quantum switch''.
\newblock \href{https://dx.doi.org/10.1103/PhysRevA.105.032217}{Phys. Rev. A {\bf 105}, 032217}~(2022).

\bibitem{unruh1976notes}
W.~G. Unruh.
\newblock ``Notes on black-hole evaporation''.
\newblock \href{https://dx.doi.org/10.1103/PhysRevD.14.870}{Phys. Rev. D {\bf 14}, 870--892}~(1976).

\bibitem{unruh1984happens}
William~G. Unruh and Robert~M. Wald.
\newblock ``What happens when an accelerating observer detects a {R}indler particle''.
\newblock \href{https://dx.doi.org/10.1103/PhysRevD.29.1047}{Phys. Rev. D {\bf 29}, 1047--1056}~(1984).

\bibitem{hawking1974black}
S.W. Hawking.
\newblock ``{Black hole explosions}''.
\newblock \href{https://dx.doi.org/10.1038/248030a0}{Nature {\bf 248}, 30--31}~(1974).

\bibitem{foo2020unruhdewitt}
Joshua Foo, Sho Onoe, and Magdalena Zych.
\newblock ``Unruh-de{W}itt detectors in quantum superpositions of trajectories''.
\newblock \href{https://dx.doi.org/10.1103/PhysRevD.102.085013}{Phys. Rev. D {\bf 102}, 085013}~(2020).

\bibitem{fooPhysRevResearch.3.043056}
Joshua Foo, Sho Onoe, Robert~B. Mann, and Magdalena Zych.
\newblock ``Thermality, causality, and the quantum-controlled {U}nruh--de{W}itt detector''.
\newblock \href{https://dx.doi.org/10.1103/PhysRevResearch.3.043056}{Phys. Rev. Research {\bf 3}, 043056}~(2021).

\bibitem{Foo_2021schrodingers}
Joshua Foo, Robert~B Mann, and Magdalena Zych.
\newblock ``Schrödinger's cat for de {S}itter spacetime''.
\newblock \href{https://dx.doi.org/10.1088/1361-6382/abf1c4}{Classical and Quantum Gravity {\bf 38}, 115010}~(2021).

\bibitem{dimic2017simulating}
Aleksandra Dimić, Marko Milivojević, Dragoljub Gočanin, Natália~S. Móller, and Časlav Brukner.
\newblock ``Simulating indefinite causal order with rindler observers''.
\newblock \href{https://dx.doi.org/10.3389/fphy.2020.525333}{Frontiers in Physics{\bf 8}}~(2020).

\bibitem{barbadoPhysRevD.102.045002}
Luis~C. Barbado, Esteban Castro-Ruiz, Luca Apadula, and Caslav Brukner.
\newblock ``Unruh effect for detectors in superposition of accelerations''.
\newblock \href{https://dx.doi.org/10.1103/PhysRevD.102.045002}{Phys. Rev. D {\bf 102}, 045002}~(2020).

\bibitem{fooPhysRevLett.129.181301}
Joshua Foo, Cemile~Senem Arabaci, Magdalena Zych, and Robert~B. Mann.
\newblock ``Quantum signatures of black hole mass superpositions''.
\newblock \href{https://dx.doi.org/10.1103/PhysRevLett.129.181301}{Phys. Rev. Lett. {\bf 129}, 181301}~(2022).

\bibitem{wood2021https://doi.org/10.48550/arxiv.2112.07860}
Carolyn~E. Wood, Harshit Verma, Fabio Costa, and Magdalena Zych.
\newblock ``Operational models of temperature superpositions''~(2021).

\bibitem{crispino_2008RevModPhys.80.787}
Lu\'{\i}s C.~B. Crispino, Atsushi Higuchi, and George E.~A. Matsas.
\newblock ``The {U}nruh effect and its applications''.
\newblock \href{https://dx.doi.org/10.1103/RevModPhys.80.787}{Rev. Mod. Phys. {\bf 80}, 787--838}~(2008).

\bibitem{sucommPhysRevD.90.084022}
Daiqin Su and T.~C. Ralph.
\newblock ``Quantum communication in the presence of a horizon''.
\newblock \href{https://dx.doi.org/10.1103/PhysRevD.90.084022}{Phys. Rev. D {\bf 90}, 084022}~(2014).

\bibitem{bekensteinPhysRevD.7.2333}
Jacob~D. Bekenstein.
\newblock ``Black holes and entropy''.
\newblock \href{https://dx.doi.org/10.1103/PhysRevD.7.2333}{Phys. Rev. D {\bf 7}, 2333--2346}~(1973).

\bibitem{Bardeen1973}
J.~M. Bardeen, B.~Carter, and S.~W. Hawking.
\newblock ``The four laws of black hole mechanics''.
\newblock \href{https://dx.doi.org/10.1007/BF01645742}{Communications in Mathematical Physics {\bf 31}, 161--170}~(1973).

\bibitem{giacomini2019}
Flaminia Giacomini, Esteban Castro-Ruiz, and Caslav Brukner.
\newblock ``{Quantum mechanics and the covariance of physical laws in quantum reference frames}''.
\newblock \href{https://dx.doi.org/https://doi.org/10.1038/s41467-018-08155-0}{Nat Commun{\bf 10}}~(2019).

\bibitem{olson2011entanglement}
S.~Jay Olson and Timothy~C. Ralph.
\newblock ``Entanglement between the future and the past in the quantum vacuum''.
\newblock \href{https://dx.doi.org/10.1103/PhysRevLett.106.110404}{Phys. Rev. Lett. {\bf 106}, 110404}~(2011).

\bibitem{suPhysRevX.9.011007}
Daiqin Su and Timothy~C. Ralph.
\newblock ``Decoherence of the radiation from an accelerated quantum source''.
\newblock \href{https://dx.doi.org/10.1103/PhysRevX.9.011007}{Phys. Rev. X {\bf 9}, 011007}~(2019).

\bibitem{fooPhysRevD.101.085006}
Joshua Foo and Timothy.~C. Ralph.
\newblock ``Continuous-variable quantum teleportation with vacuum-entangled rindler modes''.
\newblock \href{https://dx.doi.org/10.1103/PhysRevD.101.085006}{Phys. Rev. D {\bf 101}, 085006}~(2020).

\bibitem{martinetti2003diamond}
Pierre Martinetti and Carlo Rovelli.
\newblock ``Diamond's temperature: Unruh effect for bounded trajectories and thermal time hypothesis''.
\newblock \href{https://dx.doi.org/10.1088/0264-9381/20/22/015}{Classical and Quantum Gravity {\bf 20}, 4919}~(2003).

\bibitem{ida2013modular}
Daisuke Ida, Takahiro Okamoto, and Miyuki Saito.
\newblock ``Modular theory for operator algebra in a bounded region of space-time and quantum entanglement''.
\newblock \href{https://dx.doi.org/10.1093/ptep/ptt061}{Progress of Theoretical and Experimental Physics {\bf 2013}, 083E03}~(2013).
\newblock  \href{http://arxiv.org/abs/https://academic.oup.com/ptep/article-pdf/2013/8/083E03/19300976/ptt061.pdf}{arXiv:https://academic.oup.com/ptep/article-pdf/2013/8/083E03/19300976/ptt061.pdf}.

\bibitem{su2016spacetime}
Daiqin Su and T.~C. Ralph.
\newblock ``Spacetime diamonds''.
\newblock \href{https://dx.doi.org/10.1103/PhysRevD.93.044023}{Phys. Rev. D {\bf 93}, 044023}~(2016).

\bibitem{Foo_2020generating}
Joshua Foo, Sho Onoe, Magdalena Zych, and Timothy~C.\ Ralph.
\newblock ``Generating multi-partite entanglement from the quantum vacuum with a finite-lifetime mirror''.
\newblock \href{https://dx.doi.org/10.1088/1367-2630/aba1b2}{New Journal of Physics {\bf 22}, 083075}~(2020).

\bibitem{chakPhysRevD.106.045027}
A.~Chakraborty, H.~E. Camblong, and C.~R. Ord\'o\~nez.
\newblock ``Thermal effect in a causal diamond: Open quantum systems approach''.
\newblock \href{https://dx.doi.org/10.1103/PhysRevD.106.045027}{Phys. Rev. D {\bf 106}, 045027}~(2022).

\bibitem{camblongPhysRevD.110.124043}
H.~E. Camblong, A.~Chakraborty, P.~Lopez-Duque, and C.~R. Ord\'o\~nez.
\newblock ``Conformal quantum mechanics of causal diamonds: Time evolution, thermality, and instability via path integral functionals''.
\newblock \href{https://dx.doi.org/10.1103/PhysRevD.110.124043}{Phys. Rev. D {\bf 110}, 124043}~(2024).

\bibitem{camblong10.1063/5.0150349}
H.~E. Camblong, A.~Chakraborty, P.~Lopez~Duque, and C.~R. Ordóñez.
\newblock ``Spectral properties of the symmetry generators of conformal quantum mechanics: A path-integral approach''.
\newblock \href{https://dx.doi.org/10.1063/5.0150349}{Journal of Mathematical Physics {\bf 64}, 092302}~(2023).

\bibitem{jacobson10.21468/SciPostPhys.7.6.079}
Ted Jacobson and Manus~R. Visser.
\newblock ``{Gravitational thermodynamics of causal diamonds in (A)dS}''.
\newblock \href{https://dx.doi.org/10.21468/SciPostPhys.7.6.079}{SciPost Phys. {\bf 7}, 079}~(2019).

\bibitem{goodPhysRevD.102.045020}
Michael R.~R. Good, Abay Zhakenuly, and Eric~V. Linder.
\newblock ``Mirror at the edge of the universe: Reflections on an accelerated boundary correspondence with de sitter cosmology''.
\newblock \href{https://dx.doi.org/10.1103/PhysRevD.102.045020}{Phys. Rev. D {\bf 102}, 045020}~(2020).

\bibitem{GIBBONS2007317}
G.W. Gibbons and S.N. Solodukhin.
\newblock ``The geometry of small causal diamonds''.
\newblock \href{https://dx.doi.org/https://doi.org/10.1016/j.physletb.2007.03.068}{Physics Letters B {\bf 649}, 317--324}~(2007).

\bibitem{berthierePhysRevD.92.064036}
Cl\'ement Berthiere, Gary Gibbons, and Sergey~N. Solodukhin.
\newblock ``Comparison theorems for causal diamonds''.
\newblock \href{https://dx.doi.org/10.1103/PhysRevD.92.064036}{Phys. Rev. D {\bf 92}, 064036}~(2015).

\bibitem{Arzano2020}
Michele Arzano.
\newblock ``Conformal quantum mechanics of causal diamonds''.
\newblock \href{https://dx.doi.org/10.1007/JHEP05(2020)072}{Journal of High Energy Physics {\bf 2020}, 72}~(2020).

\bibitem{gradshteyn2014table}
Izrail~Solomonovich Gradshteyn and Iosif~Moiseevich Ryzhik.
\newblock ``Table of integrals, series, and products''.
\newblock \href{https://dx.doi.org/https://doi.org/10.1016/C2010-0-64839-5}{Academic press}. ~(2014).

\bibitem{LEE1986437}
T.D. Lee.
\newblock ``Are black holes black bodies?''.
\newblock \href{https://dx.doi.org/https://doi.org/10.1016/0550-3213(86)90493-1}{Nuclear Physics B {\bf 264}, 437--486}~(1986).

\bibitem{takagi10.1143/PTP.88.1}
Shin Takagi.
\newblock ``{Vacuum Noise and Stress Induced by Uniform Acceleration: Hawking-Unruh Effect in Rindler Manifold of Arbitrary Dimension}''.
\newblock \href{https://dx.doi.org/10.1143/PTP.88.1}{Progress of Theoretical Physics Supplement {\bf 88}, 1--142}~(1986).
\newblock  \href{http://arxiv.org/abs/https://academic.oup.com/ptps/article-pdf/doi/10.1143/PTP.88.1/5461184/88-1.pdf}{arXiv:https://academic.oup.com/ptps/article-pdf/doi/10.1143/PTP.88.1/5461184/88-1.pdf}.

\bibitem{bleistein1975asymptotic}
Norman Bleistein and Richard~A Handelsman.
\newblock ``Asymptotic expansions of integrals''.
\newblock Ardent Media. ~(1975).

\bibitem{birrell1984quantum}
Nicholas~David Birrell and Paul Davies.
\newblock ``Quantum fields in curved space''.
\newblock \href{https://dx.doi.org/https://doi.org/10.1017/CBO9780511622632}{Cambridge university press}. ~(1984).

\bibitem{pozasPhysRevD.94.064074}
Alejandro Pozas-Kerstjens and Eduardo Mart\'{\i}n-Mart\'{\i}nez.
\newblock ``Entanglement harvesting from the electromagnetic vacuum with hydrogenlike atoms''.
\newblock \href{https://dx.doi.org/10.1103/PhysRevD.94.064074}{Phys. Rev. D {\bf 94}, 064074}~(2016).

\bibitem{Louko_2008}
Jorma Louko and Alejandro Satz.
\newblock ``Transition rate of the {U}nruh–{D}e{W}itt detector in curved spacetime''.
\newblock \href{https://dx.doi.org/10.1088/0264-9381/25/5/055012}{Classical and Quantum Gravity {\bf 25}, 055012}~(2008).

\bibitem{stritzelbergerPhysRevD.101.036007}
Nadine Stritzelberger and Achim Kempf.
\newblock ``Coherent delocalization in the light-matter interaction''.
\newblock \href{https://dx.doi.org/10.1103/PhysRevD.101.036007}{Phys. Rev. D {\bf 101}, 036007}~(2020).

\bibitem{hendersonPhysRevLett.125.131602}
Laura~J. Henderson, Alessio Belenchia, Esteban Castro-Ruiz, Costantino Budroni, Magdalena Zych, Caslav Brukner, and Robert~B. Mann.
\newblock ``Quantum temporal superposition: The case of quantum field theory''.
\newblock \href{https://dx.doi.org/10.1103/PhysRevLett.125.131602}{Phys. Rev. Lett. {\bf 125}, 131602}~(2020).

\bibitem{tian2016detecting}
Zehua Tian, Jieci Wang, Jiliang Jing, and Andrzej Dragan.
\newblock ``Detecting the curvature of de {S}itter universe with two entangled atoms''.
\newblock \href{https://dx.doi.org/10.1038/srep35222}{Scientific Reports{\bf 6}}~(2016).

\bibitem{HUANG2017458}
Zhiming Huang and Zehua Tian.
\newblock ``Dynamics of quantum entanglement in de sitter spacetime and thermal minkowski spacetime''.
\newblock \href{https://dx.doi.org/https://doi.org/10.1016/j.nuclphysb.2017.08.014}{Nuclear Physics B {\bf 923}, 458--474}~(2017).

\bibitem{griffiths2009exact}
Jerry~B Griffiths and Ji{\v{r}}{\'\i} Podolsk{\`y}.
\newblock ``Exact space-times in {E}instein's general relativity''.
\newblock \href{https://dx.doi.org/https://doi.org/10.1017/CBO9780511635397}{Cambridge University Press}. ~(2009).

\bibitem{martinPhysRev.115.1342}
Paul~C. Martin and Julian Schwinger.
\newblock ``Theory of many-particle systems. {I}''.
\newblock \href{https://dx.doi.org/10.1103/PhysRev.115.1342}{Phys. Rev. {\bf 115}, 1342--1373}~(1959).

\bibitem{henderson2019btz}
Laura~J. Henderson, Robie~A. Hennigar, Robert~B. Mann, Alexander~R.H. Smith, and Jialin Zhang.
\newblock ``{Anti-Hawking phenomena}''.
\newblock \href{https://dx.doi.org/10.1016/j.physletb.2020.135732}{Phys. Lett. B {\bf 809}, 135732}~(2020).

\bibitem{lifschitzPhysRevD.49.1929}
Gilad Lifschytz and Miguel Ortiz.
\newblock ``Scalar field quantization on the (2+1)-dimensional black hole background''.
\newblock \href{https://dx.doi.org/10.1103/PhysRevD.49.1929}{Phys. Rev. D {\bf 49}, 1929--1943}~(1994).

\bibitem{henderson2018harvesting}
Laura~J Henderson, Robie~A Hennigar, Robert~B Mann, Alexander R~H Smith, and Jialin Zhang.
\newblock ``Harvesting entanglement from the black hole vacuum''.
\newblock \href{https://dx.doi.org/10.1088/1361-6382/aae27e}{Classical and Quantum Gravity {\bf 35}, 21LT02}~(2018).

\bibitem{paczos2022quantum}
Jerzy Paczos, Kacper Debski, Piotr~T. Grochowski, Alexander R.~H. Smith, and Andrzej Dragan.
\newblock ``Quantum time dilation in a gravitational field''.
\newblock \href{https://dx.doi.org/10.22331/q-2024-05-07-1338}{{Quantum} {\bf 8}, 1338}~(2024).

\bibitem{debski2022universality}
Kacper Debski, Piotr~T Grochowski, Rafal Demkowicz-Dobrzanski, and Andrzej Dragan.
\newblock ``Universality of quantum time dilation''.
\newblock \href{https://dx.doi.org/10.1088/1361-6382/ad4fd9}{Classical and Quantum Gravity {\bf 41}, 135014}~(2024).

\bibitem{delahamette2021falling}
Anne-Catherine de~la Hamette, Viktoria Kabel, Esteban Castro-Ruiz, and {\v{C}}aslav Brukner.
\newblock ``Quantum reference frames for an indefinite metric''.
\newblock \href{https://dx.doi.org/10.1038/s42005-023-01344-4}{Communications Physics {\bf 6}, 231}~(2023).

\bibitem{giacomini10.1116/5.0070018}
Flaminia Giacomini and Caslav Brukner.
\newblock ``{Quantum superposition of spacetimes obeys Einstein's equivalence principle}''.
\newblock \href{https://dx.doi.org/10.1116/5.0070018}{AVS Quantum Science{\bf 4}}~(2022).

\bibitem{carlip2003quantum}
Steven Carlip and Steven~Jonathan Carlip.
\newblock ``Quantum gravity in 2+ 1 dimensions''.
\newblock \href{https://dx.doi.org/https://doi.org/10.1017/CBO9780511564192}{Volume~50}.
\newblock Cambridge University Press. ~(2003).

\end{thebibliography}

\section{Appendix}

\subsection{Bogoliubov Coefficients for the Rindler and Left Rindler Modes}

\noindent The Rindler modes in the right wedge take the form (expressed in terms of the Minkowski null coordinate $V$), 
\begin{align}
    g_\omega^R (V) &= \frac{1}{\sqrt{4\pi\omega}} ( a V )^{-i\omega/a}.
\end{align}
The Klein-Gordon inner product between these modes and Minkowski modes is given by, 
\begin{align}
    \langle g_\omega^{R} (V) , u_k(V) &= 2 i \int_{0}^\infty \D V \: \frac{1}{4\pi \sqrt{\omega k}} ( a V )^{i\omega/a} (-i k) e^{-ikV} 
    \non \vt 
    \\
    &= \frac{1}{2\pi} \sqrt{\frac{k}{\omega}} \int_{0}^\infty \D V \: ( a V )^{i\omega/a}e^{-ikV} 
    \non \vt 
    \\
    &= - \frac{i e^{\pi\omega/2a}}{2\pi\sqrt{k\omega}} \Gamma(1+i\omega/a) \left( \frac{k}{a} \right)^{-i\omega/a} . 
\end{align}
For the $\beta_{k\omega}$ coefficient, we have, 
\begin{align}
    & \langle g_\omega^{R \star} (V) , u_k (V) \rangle 
    \non \vt \\
    & \:\:\: = 2 i \int_{0}^\infty \D V \: \frac{1}{4\pi \sqrt{\omega k}} ( a V )^{-i\omega/a} (-ik) e^{-ikV} 
    ,\non\vt 
    \\
    & \:\:\: = \frac{1}{2\pi} \sqrt{\frac{k}{\omega}} \frac{1}{a} \int_{0}^{\infty} \D z \: (z)^{-i\omega/a} e^{-ikz/a} . 
    \non \vt 
    \\
    & \:\:\: = - \frac{i e^{-\pi\omega/2a}}{2\pi\sqrt{k\omega}} \Gamma(1-i\omega/a) \left( \frac{k}{a} \right)^{i\omega/a}
\end{align}
To obtain the corresponding Bogoliubov coefficients for the translated wedges, one simply requires the form of the modes in these wedges. This is given by, 
\begin{align}
    g_\omega^R (V) &= \frac{1}{\sqrt{4\pi\omega}} (aV + s)^{-i\omega/a},
\end{align}
where the translation is $-s/a$ in the null coordinate. One finds after a similar calculation, that 
\begin{align}
    \alpha_{k\omega}^{R\prime} &= e^{iks/a} \alpha_{k\omega}^R, 
    \vt \\
    \beta_{k\omega}^{R\prime} &= e^{iks/a} \beta_{k\omega}^R , \vt 
\end{align}
as stated in the main text. Similarly for the left Rindler modes, we have,
\begin{align}
    g_\omega^L(V) &= \frac{1}{\sqrt{4\pi\omega}} (-aV)^{i\omega/a}, 
\end{align}
which gives, a Klein-Gordon inner product of the form,
\begin{align}
    \langle g_\omega^{L} , u_k(V) \rangle &= 2 i \int_{-\infty}^{0} \D V \: \frac{1}{4\pi\sqrt{\omega k}} (- a V )^{-i\omega/a} (-ik) e^{-ikV} 
    \non 
    \\ 
    &= \frac{i e^{\pi\omega/2a} }{2\pi\sqrt{\omega k}} \left(\frac{k}{a} \right)^{i\omega/a} \Gamma(1-i\omega/a)
\end{align}
while for the $\beta_{k\omega}$ coefficients,
\begin{align}
    \langle g_\omega^{L\star},u_k(V) \rangle 
    &= \frac{1}{2\pi a} e^{iks/a} \sqrt{\frac{k}{\omega}} \int_0^\infty \D z \: (z)^{i\omega/a} e^{ikz/a} 
    \non \vt
    \\
    &= \frac{i e^{-\pi\omega/2a}}{2\pi\sqrt{\omega k}} e^{iks/a} \left( \frac{k}{a} \right)^{-i\omega/a} \Gamma(1+i\omega/a) .
\end{align}
Note that the $\beta$ Bogoliubov coefficients correspond with those computed in Ref.\ \cite{crispino_2008RevModPhys.80.787} while the $\alpha$ coefficients differ by a global phase. This is due to a trick used in Ref.\ \cite{crispino_2008RevModPhys.80.787} to compute $\alpha_{k\omega}^{R,L}$ from the Minkowski decomposition of the Rindler modes, rather than explicitly using the Klein-Gordon inner product. Of course, this phase is irrelevant for computing observable quantities. As with the right Rindler modes, it is straightforward to show that the Bogoliubov coefficients for the shifted wedges is given by, 
\begin{align}
    \alpha_{k\omega}^{L\prime} &= e^{iks/a} \alpha_{k\omega}^L , \vt 
    \\
    \beta_{k\omega}^{L\prime} &= e^{iks/a} \beta_{k\omega}^L . \vt 
\end{align}

\subsection{Overlap of Shifted Right Rindler Modes}

\noindent We have, 
\begin{align}
    \beta_{k\omega}^R &= - \frac{i e^{-\pi\omega/2a}}{2\pi\sqrt{k\omega}} \Gamma(1-i\omega/a) \left( \frac{k}{a} \right)^{i\omega/a}
\end{align}
and the relation $\beta_{k\omega}^{R\prime} = e^{iks/a} \beta_{k\omega}^R$. The relevant cross-term is given by, 
\begin{align}
    I &= \int\D k \: \beta_{k\omega}^{R\star} \beta_{k\omega'}^{R\prime} = \Lambda(\Omega, \Omega') \int\D k \: \frac{1}{k} k^{-i(\omega-\omega')/a} e^{iks/a} 
\end{align}
To evaluate the integral, we use $e^{iks/a} = \cos(ks/a) + i \sin(ks/a)$ which explicitly gives \begin{align} 
    I &= \Lambda (\Omega , \Omega' ) \left( \frac{s}{a} \right)^{i(\Omega - \Omega')} e^{\pi(\Omega- \Omega')} \Gamma\Big[ -i (\Omega - \Omega' ) \Big] 
    \nonumber 
\end{align}
where 
\begin{align}
    \Lambda(\Omega,\Omega') &= \frac{e^{-\pi(\Omega+\Omega')/2}}{2\pi \sqrt{\Omega\Omega'}} \Gamma(1+i\Omega) \Gamma(1-i\Omega') \left( \frac{1}{a} \right)^{-i(\Omega-\Omega') }. 
    \nonumber 
\end{align}
The calculation of the overlap between the right Rindler modes and the shifted left modes follows analogously. 

\newpage 

\begin{widetext} 

\subsection{Derivation of the Diamond Temperature Using Bogoliubov Coefficients}

\noindent Using the integral form of the Bogoliubov coefficients, the vacuum expectation value of the particle number is given by
\begin{align}
    N_{\omega\omega'} &= \int_0^\infty\D k \: \beta_{\omega k}^{(0)} \beta_{\omega'k}^{(0)} ,
    \non 
    \\
    &= \frac{1}{\pi^2 a \sqrt{\Omega\Omega'}} \int_{-1}^{+1} \D s \int_{-1}^{+1} \D s' \left( \frac{1+s}{1-s} \right)^{-i\Omega} \left( \frac{1 + s'}{1-s'}\right)^{i\Omega'} \int_0^\infty\D \kappa \: \kappa e^{-2i\kappa(s-s')},
    \non 
    \\
    &= - \frac{1}{4\pi^2 a \sqrt{\Omega\Omega'}} \int_{-1}^{+1} \D s \int_{-1}^{+1} \D s' \frac{1}{(s-s'-i\varepsilon)^2} \left( \frac{1+s}{1 - s} \right)^{-i\Omega} \left( \frac{1 + s'}{1 -s'} \right)^{i\Omega'} ,
\end{align}
where in the last equality, we have used the result 
\begin{align}
    \int_0^\infty \D \kappa \: \kappa e^{-i\kappa z} &= - \frac{1}{(z - i\varepsilon)^2}.
\end{align}
Defining new integration variables, 
\begin{align}
    t &= \frac{1}{2} \ln \left( \frac{1 + s}{1 - s} \right) \:\:\:\:\: \Longleftrightarrow \:\:\:\:\: s = \tanh t ,
\end{align}
yields 
\begin{align}
    N_{\omega\omega'} &= - \frac{1}{4\pi^2 a\sqrt{\Omega\Omega'}} \infint\D t \infint\D t' \frac{e^{-2i(\Omega t- \Omega't')}}{\sinh^2(t-t'-i\varepsilon)}.
\end{align}
Using the sum-difference change of variables 
\begin{align}
    p &= t + t', \:\:\:\:\: q = t-t',
\end{align}
we find that 
\begin{align}
    N_{\omega\omega'} &= - \frac{1}{8\pi^2a \sqrt{\Omega\Omega'}} \infint\D p \: e^{-i(\Omega - \Omega') p} \infint\D q \: \frac{e^{-i(\Omega + \Omega')q}}{\sinh^2(q-i\varepsilon)} ,
    \non 
    \\
    &= - \frac{1}{4\pi^2a \Omega} \delta(\Omega - \Omega' ) \infint\D q \: \frac{e^{-2i\Omega q}}{\sinh^2(q - i\varepsilon)},
    \non 
    \\
    &= \frac{\delta ( \omega - \omega') }{e^{2\pi\omega/a} - 1} \vphantom{\infint}. 
\end{align}
This is a thermal distribution at the temperature $T_D = a(2\pi)^{-1}$ as desired.

\subsection{Overlap of Shifted Diamond Modes}

\noindent The Bogoliubov coefficients are given by
\begin{align}
    \beta_{\omega k}^{(0)} &= - \frac{2}{a} \frac{\sqrt{\Omega\kappa}}{\sinh(\pi\Omega)} e^{-2i\kappa} M ( 1 + i \Omega, 2 , 4i \kappa ) ,
\end{align}
which is related to the shifted diamond modes via the relation $\beta_{\omega k}^{(n)} = e^{4i\kappa n} \beta_{\omega k}^{(0)}$. The overlap is given by, 
\begin{align}
    \int\D k \: \beta_{\omega k}^{(0)\star} \beta_{\omega'k}^{(n)} &= \frac{4}{a^3}\frac{ \sqrt{\Omega\Omega'}}{\sinh(\pi\Omega) \sinh(\pi\Omega')} \int\D k \: k e^{2ik(1+2n)/a} M(1 -i \Omega, 2, -4i k/a ) M ( 1 + i\Omega', 2, 4i k/a )  .
\end{align}
Here we can utilise Eq.\ (7.622) of \cite{gradshteyn2014table}, 
\begin{align}
    & \int_0^\infty \D k \: e^{-sk} k^{c-1} M ( b,c,k) M(\alpha,c,\lambda k) 
    \Gamma(c) (s-1)^{-b} (s-\lambda)^{b+\alpha - c} F\Big[ b,\alpha;c;\lambda(s-1)^{-1} (s-\lambda)^{-1} \Big] ,
\end{align}
to express the $k$ integral as, 
\begin{align}
    I_\mathrm{K} &= ( \alpha )^{-i(\Omega - \Omega')} ( \alpha -1  )^{i\Omega-1} (\alpha+1)^{-1-i\Omega'} F \Big[ 1- i \Omega, 1 + i \Omega';2;-(\alpha-1)^{-1}(-\alpha+1)^{-1} \Big],
\end{align}
which is the result stated in the main text. 

\end{widetext}

\subsection{Unruh-deWitt Detector Response for a Superposition of Angular Positions in the BTZ Spacetime}
\noindent The minimal Unruh-deWitt coupling is described by interaction Hamiltonian, 
\begin{align}
    H_\mathrm{int.} &= \lambda \eta(\tau) \underbrace{ ( | e \rangle \langle g | e^{i\Omega \tau} + \mathrm{H.c} )}_{\hat{\sigma}(\tau)} \otimes \hat{\phi}(\mathbf{X}),
\end{align}
where $\eta(\tau)$ is a $\tau$-dependent switching function, $\hat{\sigma}(\tau) $ is a ladder operator between the detector's internal levels $| g \rangle, | e \rangle$ separated by gap $\Omega$, and $\hat{\phi}(\mathbf{X})$ is a massless scalar field pulled back to the worldline $\mathbf{X}$ of the detector's worldline. To leading order in the small coupling constant $\lambda$, the response of the detector is,
\begin{align}
    F &= \infint\D s \: e^{-s^2/4\sigma^2} e^{-i\Omega s}W(s) .
    \label{eq91}
\end{align}
where we absorbed a factor of $\lambda^2 \sigma$. $W(s) \equiv \langle \Phi | \hat{\phi}(\mathbf{X} ) \hat{\phi}(\mathbf{X}' ) | \Phi\rangle$ is the Wightman function evaluated at two times along the detector's worldline given that the field is in the state $| \Phi \rangle$. Introducing a control DoF $i$, whose states $| i \rangle$ dictate the angular position of the detector outside the black hole. The Hamiltonian is now modified to read, 
\begin{align}
    H_\mathrm{int.} &= \lambda \sum_i \eta(\tau_i) \hat{\sigma}(\tau_i) \otimes \hat{\phi}(\mathbf{X}_i) \otimes | i \rangle \langle i | , 
\end{align}
where $\tau_i$ denotes the proper time of the detector along the worldline $\mathbf{X}_i$. Since we are considering the detector situated at the same radial distance from the black hole, it is sufficient to drop the subscript $i$ since the proper time will be equal along each of the worldlines. Now, the response is modified accordingly, given by, 
\begin{align}
    F &= \frac{1}{2} \infint \D s \: e^{-s^2/4\sigma^2} e^{-i\Omega s} \Big[ W_A(s) + W_{AB}(s) \Big] ,
    \label{eq116}
\end{align}
where $W_A(s) = W_B(s)$ is the Wightman function for each of the individual paths of the detector, while $W_{AB}(s) \equiv \langle \Phi | \hat{\phi}(\mathbf{X}_A) \hat{\phi}(\mathbf{X}_B') | \Phi \rangle$ is a two-point correlator evaluated with respect to the two worldlines in superposition. The vacuum Wightman function for the BTZ black hole, assuming transparent boundary conditions \cite{carlip2003quantum}, is 
\begin{widetext} 
\begin{align}
    W_{A}(s) \equiv W_B(s) &= \frac{1}{4\pi l \sqrt{2}} \sum_{n=-\infty}^{+\infty} \frac{\sqrt{M}}{ \sqrt{f(r)}} \frac{1}{\sqrt{\frac{r^2}{l^2f(r)} \cosh\left[ 2\pi \sqrt{M}n \right] - \frac{M}{f(r)} - \cosh \left[ \frac{\sqrt{M}s}{l\sqrt{f(r)}} \right] }}.
    \label{eq115}
\end{align}
\end{widetext} 

\newpage 

\noindent Inserting Eq.\ (\ref{eq115}) into Eq.\ (\ref{eq91}) gives, local contributions to the response function (see Appendix of \cite{henderson2019btz}),
\begin{align}
    F ( \Omega) &=  \frac{\sqrt{\pi}/2}{e^{\Omega/T}+1} \sum_{n=-\infty}^{+\infty} P_{-\frac{1}{2} + \frac{i\Omega}{2\pi T} } (\cosh\alpha_n^-) ,
    \label{eq118}
    \vspace{20pt}
\end{align}
as stated in the main text. Meanwhile, the nonlocal Wightman function modifies Eq.\ (\ref{eq115}) by the inclusion of the angular separation $\delta\phi$, 
\begin{widetext}
\begin{align}
    W_{AB}(s) &= \frac{1}{4\pi l \sqrt{2}} \sum_{n=-\infty}^{+\infty} \frac{\sqrt{M}}{f(r)} \frac{1}{\sqrt{\frac{r^2}{l^2f(r)} \cosh(\sqrt{M}(\delta \phi - 2\pi n ) - \frac{M}{f(r)} - \cosh \left[ \frac{\sqrt{M}s}{l\sqrt{f(r)}} \right]}} .
\end{align}
\end{widetext} 
This gives a nonlocal contribution to the response (i.e.\ the interference term) of the same form as Eq.\ (\ref{eq118}) with the replacement, 
\begin{align}
    \alpha_n^- \to \beta_n^- = ( 1 + X ) \cosh(\sqrt{M}(\delta \phi - 2\pi n ))  - X 
\end{align}
as stated in the main text.

\end{document}